\documentclass[3p,preprint,12pt]{elsarticle}

\usepackage{amssymb}
\usepackage{amsmath}
\usepackage[brazil]{babel}
\usepackage[T1]{fontenc}
\usepackage[utf8]{inputenc}

\journal{Fundamental Plasma Physics}

\begin{document}

\begin{frontmatter}



\title{Confining and escaping magnetic field lines in Tokamaks:\\ 
Analysis via symplectic maps}


\author[a]{Matheus S. Palmero}
\author[a]{Iber\^e L. Caldas}

\affiliation[a]{organization={Institute of Physics, University of S\~ao Paulo},
            addressline={Rua do Mat\~ao 1371}, 
            city={S\~ao Paulo},
            postcode={05508-090}, 
            state={S\~ao Paulo},
            country={Brazil}}

\begin{abstract}
In magnetically confined plasma, it is possible to qualitatively describe the magnetic field configuration via phase spaces of suitable symplectic maps. These phase spaces are of mixed type, where chaos coexists with regular motion, and the complete understanding of the chaotic transport is a challenge that, when overcome, may provide further knowledge into the behaviour of confined fusion plasma. This work presents two numerical investigations into characteristics of mixed phase spaces which model distinct magnetic configurations in tokamaks under different perturbation regimes. The first approach relies on a recurrence-based analysis of ensembles of chaotic trajectories to detect open field lines that widely differ from the average. The second focuses on the transient dynamical behaviour of field lines before they escape the systems. These two methods provide insights into the influence of stickiness and invariant manifolds on the evolution of chaotic trajectories, improving our understanding of how these features affect the chaotic transport and diffusion properties in mixed phase spaces. These theoretical and numerical approaches may enhance our comprehension of confined plasma behaviour and plasma-wall interactions.
\end{abstract}

\begin{graphicalabstract}
\centering
\includegraphics{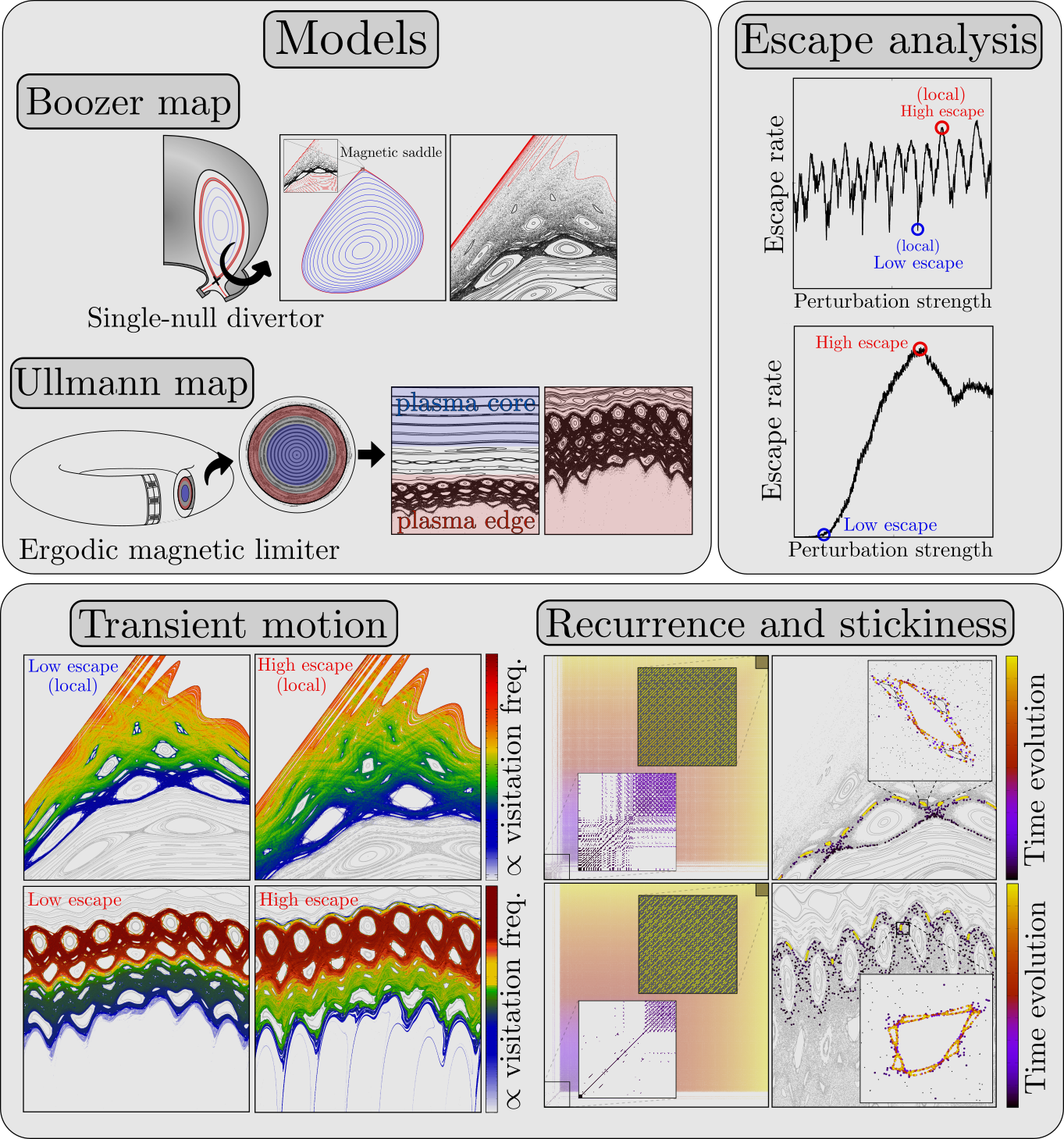}
\end{graphicalabstract}



\end{frontmatter}


\section{Introduction}
\label{intro}
Decades of research have been devoted to investigate Hamiltonian systems under small periodic perturbations, primarily due to their complex and rich dynamical properties in both chaotic and periodic motion \cite{Lichtenberg1992,Ott2002,Meiss2015}. In particular cases, these systems can be described by symplectic two-dimensional non-linear maps, which often replicate the general dynamical behaviour of higher dimensional dynamical systems in many areas of science \cite{Henon1976,Meiss1992,Leonel2004}. Our current research focus on the application in plasma dynamics, specifically modelling the configuration of the magnetic field in tokamaks.

Tokamaks \cite{Wesson2004} are promising machines of magnetically confined plasma to achieve thermonuclear fusion. To efficiently produce energy, important problems are still under investigation, expected to have further definitive results over the next few years in experiences on the ITER \cite{Horton2015} tokamak.  

In modern tokamaks, like ITER, one of the most investigated subjects for enhancing the magnetic confinement is the topology of the magnetic field \cite{Robinson1993, Connor1993}, the so called \emph{magnetic configuration} of the system. These configurations are constantly perturbed by natural or induced oscillations to control the plasma \cite{Horton2014, Morrison1998}. In the plasma core, the magnetic field lines often form stable toroidal magnetic surfaces. However, resonant magnetic perturbations, especially on the plasma edge, can break those surfaces forming unstable regions with open field lines that escape the confinement, dragging particles from the plasma to the inner wall of the tokamak chamber. This process, when not controlled, may damage the machine.

In the context of nonlinear dynamics, open magnetic field lines are related to chaotic orbits \cite{Morrison2000} wandering through the chaotic sea in the phase space. Modelling magnetic field lines by Hamiltonian maps allows the investigation of magnetic configurations via the phase spaces of the models. Moreover, open field lines in tokamaks can be controlled by applying external electric currents or magnetic devices that alters the configuration on the plasma edge, altering the phase space of the model as well.

Different experimental evidence \cite{Finken2006,Evans2015} and theoretical models \cite{Kroetz2012,Ciro2016} suggest that the control of the magnetic field lines acts on the transport of particles on the plasma edge. This transport is considered to be \emph{anomalous} \cite{Negrete2006} because it is different from what is proposed by normal diffusion. Moreover, modelling the anomalous transport on the phase space of the field lines maps is a theoretical/numerical approach that helps on the understanding of the confined plasma behaviour and plasma-wall interactions. 

Near-integrable Hamiltonian systems presents area-preserving phase spaces that are composed by a local and/or global chaotic sea along with KAM islands of periodic dynamics \cite{Zaslavsky2002}. These are called \emph{mixed phase spaces}, where a chaotic orbit evolved from initial conditions (ICs) at the chaotic sea, may experience different dynamical behaviours in a given maximum iteration time, or until it escapes the system. In that sense, a \emph{transient dynamics} is the evolution of a single orbit, or an ensemble of orbits, until escape the systems.

Particularly for mixed phase spaces, there exist regions in which chaotic orbits spend an expressive amount of time experiencing successive dynamical traps. The trajectory, once free to explore all chaotic regions of the phase space, is now temporally confined in a quasi-periodic motion in the vicinity around the stability islands. This is the well-known phenomenon of \emph{stickiness} \cite{Altmann2006,Lai2011,Contopoulos2010}, that affects transport and statistical properties of chaotic orbits. In that sense, prior knowledge if a given trajectory will or will not experience stickiness is important. 

In addition to stickiness, mixed phase spaces of non-linear symplectic maps present complicated intertwined underlying structures that strongly affect the transport properties of the system. These structures are known as \emph{invariant manifolds} and determining how these structures are spatially organised is essential to comprehend how different regions of a given phase space are linked or severed. Invariant manifolds often act as transport barriers, partial transport barriers and transport channels \cite{Wiggins2003,Alligood2012}. Indeed, Borgogno {\it et al} \cite{Borgogno2008} investigate influences of stable and unstable manifolds and the presence of stickiness in the context of magnetic configurations generated by nonlinear reconnection.  

In this work, we combine two numerical investigations on the aforementioned characteristics of mixed phase spaces; stickiness detection and organisation of invariant manifolds, considering the context of magnetic field lines in tokamaks under different perturbation regimes. The first approach, designed to study stickiness, is based on recurrence-based analysis and detection of stickiness in large ensembles of chaotic trajectories \cite{Palmero2022}. The second, developed to investigate the transient behaviour of open field lines before escaping, shows the influence of invariant manifolds in the general dynamical behaviour \cite{deOliveira2022}. These numerical investigations were carried out on selected symplectic maps for tokamaks under two different configurations; The \emph{Single-null divertor map}, or Boozer map \cite{Punjabi1992}, a phenomenological model that describes the magnetic configuration of a tokamak equipped with a poloidal divertor and; The \emph{Ergodic magnetic limiter map}, or Ullmann map \cite{Ullmann2000}, a parametric adjustable map that describes the magnetic field lines of a tokamak assembled with an ergodic magnetic limiter.

Our results illustrate the differences between the behaviour of field lines in both models while considering induced magnetic configurations that either enhance or restrain the escaping field lines. The first analysis identifies trajectories that widely differs from the average chaotic behaviour, specifically detecting the stickiness phenomenon, which can be related to additional confinement regions in the nearest surroundings of magnetic islands in the plasma edge. The second shows how the spatial organisation of invariant manifolds creates fitting transport channels for the open field lines, influencing the average dynamical evolution in the correspondent magnetic configuration. These analyses may, ultimately, assist in selecting optimal experimental parameters to achieve specific goals in tokamak discharges. 

This manuscript is organised as follows: Next section is devoted to the description of the models and a escape analysis that provides a general methodology for comparing and studying different magnetic configurations; In the third section, we present a brief explanation to the numerical methods, showing our results and observations; Finally, we draw our conclusions in the final section.

\section{The models}
\label{models}
In this section we present and discuss the two selected symplectic maps that model the magnetic configuration of tokamaks with different setups. 

\subsection{Single-null divertor map}
\label{models:boozer}

\begin{figure}[h!]
\centering
\includegraphics[scale=0.9]{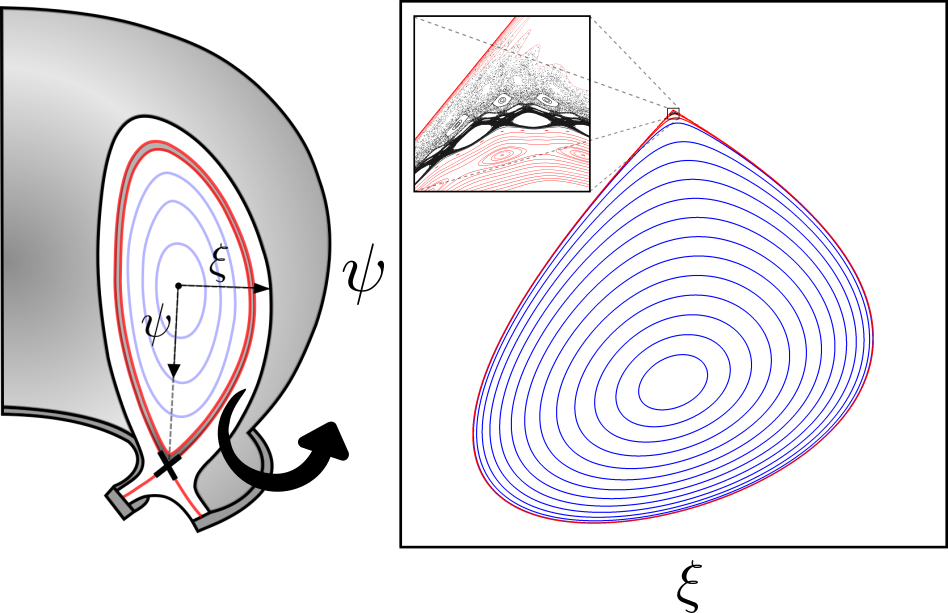}
\caption{Schematics of a poloidal section of a divertor tokamak, showing the closed magnetic field lines (light blue), the region around the magnetic separatrix (grey region between the red lines), magnetic saddle (black cross) and the general rectangular coordinates $(\xi,\psi)$. The inset shows the region around the magnetic saddle.}
\label{fig:BM_scheme}
\end{figure}

\begin{figure}[b!]
\centering
\includegraphics[scale=0.75]{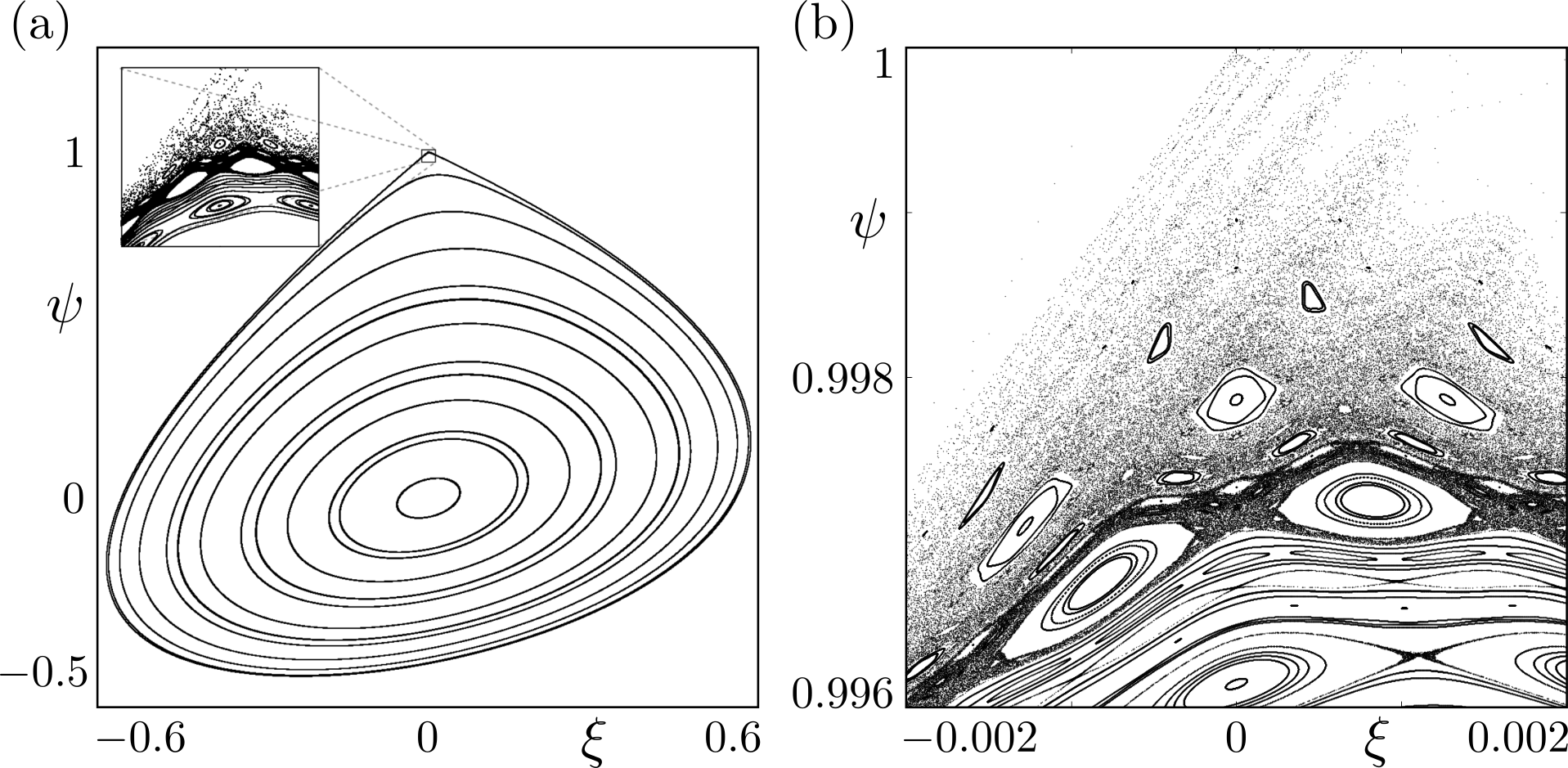}
\caption{Characteristic phase space of the BM considering $k = 0.6$. (a) Full phase space with the inset showing the region around the magnetic saddle; (b) The inset amplified, showing the mixed chaotic-periodic states present in this region.}
\label{fig:BM_phase_space}
\end{figure}

The single-null divertor map, also known as the \emph{Boozer map} (BM), was proposed by Punjabi, Verma and Boozer \cite{Punjabi1992} as a phenomenological model for the magnetic field lines of a tokamak equipped with poloidal divertors. Divertors are external devices placed at a poloidal section of the tokamak, designed to exhaust unwanted particles from the plasma to maintain the fusion reaction inside the tokamak. 

Technically, the divertor induces a magnetic configuration with a saddle point (x-point) known as the \emph{magnetic saddle}. This setup allows particles to follow the field lines towards exit points precisely placed near the divertor targets. Due to perturbations in the magnetic field, a chaotic layer is formed around the saddle, allowing the open field lines to escape through the x-points, striking the divertor target. The striking points are commonly referred to as \emph{magnetic footprints}, which form specific patterns explored in different works in the literature \cite{Punjabi1997}. Indeed, experimental evidence \cite{Ciro2016} based on heat patterns at the divertor target, suggests that ions from the plasma may follow open magnetic field lines, striking the target and forming specific heat signatures comparable to theoretical magnetic footprints.

The behaviour of field lines around the magnetic saddle is studied via the symplectic separatrix map $T_{\text{BM}}$ given by the following equations
\begin{equation}
T_{\text{BM}}:\left\{\begin{array}{ll}
\xi_{n+1}=\xi_n-k\psi_n(1-\psi_n)\\
\psi_{n+1}=\psi_n+k\xi_{n+1}
\end{array}
\right.~,
\label{eq:bm_map}
\end{equation}where the pair $(\xi, \psi)$ are generic rectangular coordinates over a poloidal section surface, as depicted in Fig.~\ref{fig:BM_scheme}; Note that $\psi$ is positive from the centre of the poloidal section in the direction of the x-point. 

The control parameter $k$ is related to the amplitude of toroidal asymmetries that perturb the magnetic field configuration. Indeed, there are a few works in the literature that relates numerical values of $k$ to the safety factor calculated at the plasma edge. Accordingly to these works, $k \approx 0.6$ is a fair value to simulate the diverted magnetic field configuration while considering, specifically, large tokamaks like ITER \cite{Punjabi1997}. 

The characteristic phase space of the model is shown in Fig.\ \ref{fig:BM_phase_space}. In particular, panel (b) shows the amplified region around the magnetic saddle located at $(\xi^{\star},\psi^{\star}) = (0,1)$, where the separatrix chaotic layer, embedded with several highly-periodic island chains, is clearly visible. The chaotic layer is essentially composed of open field lines that will, eventually, escape through the x-point, hitting the divertor target. In that sense, the escape condition for these chaotic trajectories is satisfied when $\psi_n \leq \psi_{\text{target}} \leq \psi_{n+1}$; We consider $\psi_{\text{target}} = 1.0$ for our numerical simulations in Sec. \ref{escape} and \ref{results}.

Although simple, the BM is a very suitable model to perform extensive numerical simulations that might improve our understanding of the complex behaviour of the magnetic field lines around magnetic saddles in divertor tokamaks. 

\subsection{Ergodic magnetic limiter map}
\label{models:ullmann}

\begin{figure}[h!]
\centering
\includegraphics[scale=0.65]{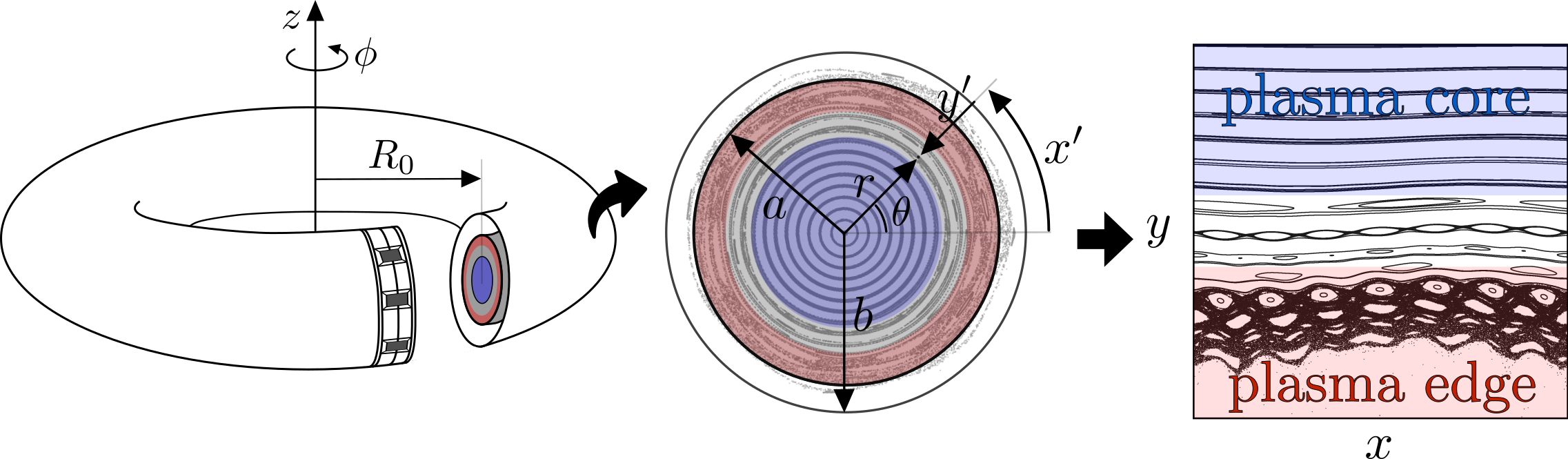}
\caption{Schematic example of the modelling process that begins at the torus (representing the tokamak), passing by the poloidal section near the limiter (ring placed in a section of the torus), showing also the calculated magnetic field lines (grey in the background), arriving at the rectangular coordinates $(x,y)$ drawing the full phase space of the model. The colour blue represents regions closer to the plasma core and the colour red regions around the plasma edge.}
\label{fig:UM_scheme}
\end{figure}

The ergodic magnetic limiter map, or the \emph{Ullmann map} (UM), was proposed as a symplectic two-dimensional non-linear map that models the magnetic field lines of a tokamak assembled with an ergodic limiter \cite{Ullmann2000}. Inside the tokamak, in the plasma core, the magnetic field is strong and stable enough for the duration of a typical discharge. However, on the plasma edge, closer to the inner walls of the machine, the magnetic field lines are often perturbed, forming regions of strong instabilities. In many cases, to either control or change the magnetic configuration in this outer region, the tokamak is assembled with devices placed at the border of the machine. This is the case of the ergodic magnetic limiter which is, basically, an outer ring composed of several helical coils that perturb the field lines at the plasma edge.  

In the practical sense, there are a few features that make the UM a fitting model for our analysis: {\it (i)} Since it is a symplectic map, it can be derived from suitable generating functions that include the appropriate periodic perturbation; {\it (ii)} The profile of the safety factor $q(r)$ is freely adjustable for a given tokamak discharge, allowing also an analysis considering non-monotonic profiles \cite{Osorio2021}; {\it(iii)} The parameters of the model are directly linked to experimental parameters of a tokamak, such as the intensity of the toroidal magnetic field $B_0$, large radius $R_0$, small radius $b$, and radius of the plasma column $a$ \cite{}. Figure \ref{fig:UM_scheme} displays a schematic example of the modelling process, along with the aforementioned parameters.

The full derivation of the model is outlined in \cite{Ullmann2000}. Essentially, the complete model is a composition of two maps $T^0_{\text{UM}} \circ T^1_{\text{UM}} (x_n,y_n) = (x_{n+1},y_{n+1})$. The first part $T^0_{\text{UM}}$ is the equilibrium dynamics with a toroidal correction to the field line equations proposed by Ullmann, resulting on the following map 

\begin{equation}
T^0_{\text{UM}}:\left\{\begin{array}{ll}
y_{n+1}^{*}=1-\frac{(1 - y_n)}{1-a_1\sin(x_n)}\\
x_{n+1}^{*}=x_n+\frac{2\pi}{q^{0}(y_{n+1}^{*})}+a_1\cos(x_n)\\
\end{array}
\right.~,
\label{eq:um_eq}
\end{equation}where $q^{0}(y)$ is the safety factor, obtained from the poloidal magnetic field $B^0_\theta(r)$, calculated at the new dimensionless coordinate $y$.

\begin{figure}[b!]
\centering
\includegraphics[scale=0.8]{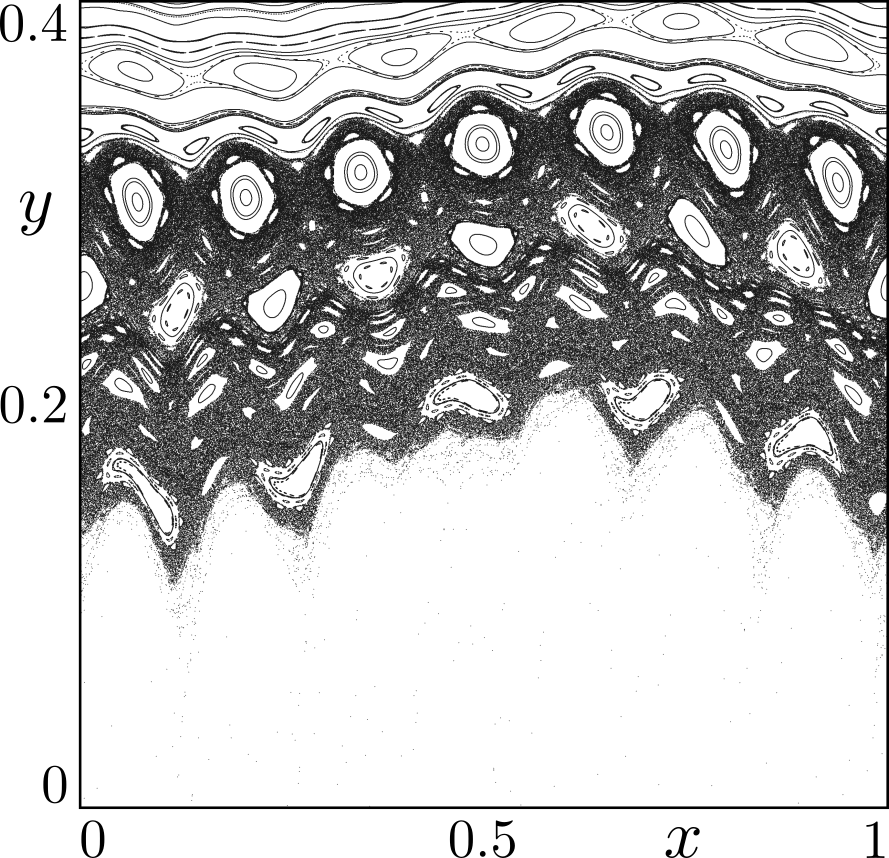}
\caption{Characteristic phase space of the UM considering $m = 7$, $q^0(a) = 5.0$ and $\delta B = 1.5 \%$ ($\delta I \approx 0.45\%$).}
\label{fig:UM_phase_space}
\end{figure}

The second part $T^{\text{pert}}_{\text{UM}}$ is considering the periodic perturbation caused by the ergodic magnetic limiter given by the following equations

\begin{equation}
T^1_{\text{UM}}:\left\{\begin{array}{ll}
y_{n}=y_{n+1}^{*}+\frac{m}{m-1}C(1-y_n)^{m-1}\sin(m x_n^{*})\\
x_{n+1}=x_n^{*}-C(1-y_{n})^{m-2}\cos(m x_n^{*})\\
\end{array}
\right.~,
\label{eq:um_pert}
\end{equation}where the dimensionless constant $C$ arrange all main parameters of the model, relating them in the following way

\begin{equation}
    C = \frac{2 \pi}{q^0(a)}\left(\frac{b}{a}\right)^{m-2} \frac{B^1(a)}{B^0_\theta(a)} = \frac{4 m \pi}{q^0(a)} \left(\frac{a}{b}\right)^2 \frac{I_h}{I_p}~.
    \label{eq:um_c}
\end{equation}Defining $\delta B = B^1(a) / B^0_\theta(a)$ and $\delta I = I_h / I_p$, we finally reach the proper control parameter of the model: $\delta B$ defined as the relative perturbation of the poloidal magnetic field or; $\delta I$ as the relative current factor. In practical terms, $\delta B = 1.0\%$ means that the intensity of the magnetic field caused by the limiter is $1.0\%$ of the intensity of the poloidal magnetic field at the plasma edge or, $\delta I = 0.5\%$ means that the electric current of the limiter is $0.5\%$ of the plasma current. 

With that, all parameters of the model can be set for a specific tokamak and, the iteration of the maps $T^0_{\text{UM}} \circ T^1_{\text{UM}}$ provide different magnetic field configurations considering different values of $\delta B$ or $\delta I$. In our numerical simulations, we use the parameters of the TCABR, the tokamak of the Physics Institute, University of S\~ao Paulo, given by the following table:   

\begin{table}[h!]
   \caption{Main parameters of the TCABR \cite{Elton2002}} 
   \label{tcabr}
   \small 
   \centering 
   \begin{tabular}{lccr} 
   \hline\hline   
   \textbf{Parameter} & \textbf{Symbol} & \textbf{Value} & \\ 
   \hline
   Larger radius & $R_0$ & $0,615$ m \\
   Minor radius & $b$ & $0,21$ m \\
   Plasma column radius & $a$ & $0,18$ m \\
   Toroidal magnetic field & $B_0$ & $1,07$ T \\
   Plasma current (equilibrium) & $I_p$ & $< 100$ kA \\
   Safety factor (equilibrium at $r=a$) & $q^0(a)$ & $\approx 5.0$ \\
   \hline
   \end{tabular}
\end{table}The full phase space of the model was presented in the last panel of Fig.\ \ref{fig:UM_scheme} however, from now on, we focus only on the region of the plasma edge $0.0 < y < 0.4$. Figure \ref{fig:UM_phase_space} displays the characteristic phase space of the UM considering the perturbation mode $m = 7$, the safety factor $q^0(a) = 5.0$ and $\delta B = 1.5 \%$ ($\delta I \approx 0.45\%$).

The chaotic region, depicted by the extended connected black sea around the periodic islands in white, present in phase space drawn in Fig.\ \ref{fig:UM_phase_space} will be an essential framework for all phase space analyses in next sections. It's worth noting that the open field lines, that compose the aforementioned chaotic sea, can eventually escape hitting the inner wall of the tokamak at $y=0$. Therefore, the model's escape condition is satisfied when $y_n < 0 < y_{n+1}$.

Escaping field lines, featured in both BM and UM models, are the main focus of the methodology proposed in the next section.

\section{Escape analysis}
\label{escape}
The methodology outlined here was developed to address one fundamental practical question: Which values of control parameters should be selected for all numerical simulations of the models?

Our approach to addressing this question is based on the escape rate $ER$, which is defined as the proportion of escaping trajectories ${\bf x}^e$ that correspond to the escaping magnetic field lines, relative to the total number $M$ of initial conditions (ICs) provided to the models;$ER = {\bf x}^e / M$. By defining $ER$, we can analyse it as a function of the control parameters, $k$ (for the BM) and $\delta B$ (for the UM). To do so, we establish a suitable range of parameter values that preserves the key characteristic features of the phase spaces. The parameter ranges are defined as follows:

\begin{equation}
    \begin{aligned}
    k\in\frac{[0.53, 0.63]}{L}~\text{and}~
    \delta B \in \frac{[1.2\%, 2.2\%]}{L}~,
    \end{aligned}
    \label{eq:method_ranges}
\end{equation}where $L$ is the number of parameters between the considered range. 

\begin{figure}[h!]
\centering
\includegraphics[scale=0.59]{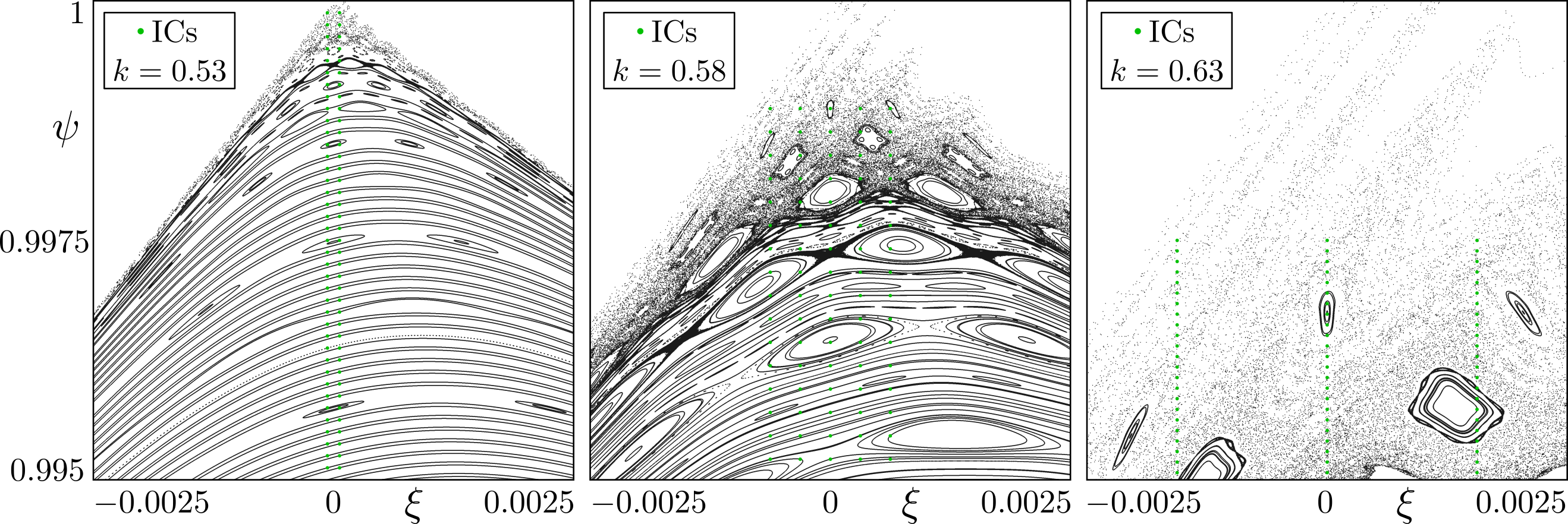}
\caption{Phase spaces of the BM considering $k=0.53$ (left), $k=0.58$ (centre) and $k=0.63$ (right). The selected ICs are represented by the green points and their evolution was up to $5 \times 10^5$ iterations of the map $T_{\text{BM}}$. All three phase spaces are drawn considering only the region $\xi \in [-0.0025, 0.0025]$ and $\psi \in [0.995, 1.0]$.}
\label{fig:BM_3ps}
\end{figure}

\begin{figure}[h!]
\centering
\includegraphics[scale=0.6]{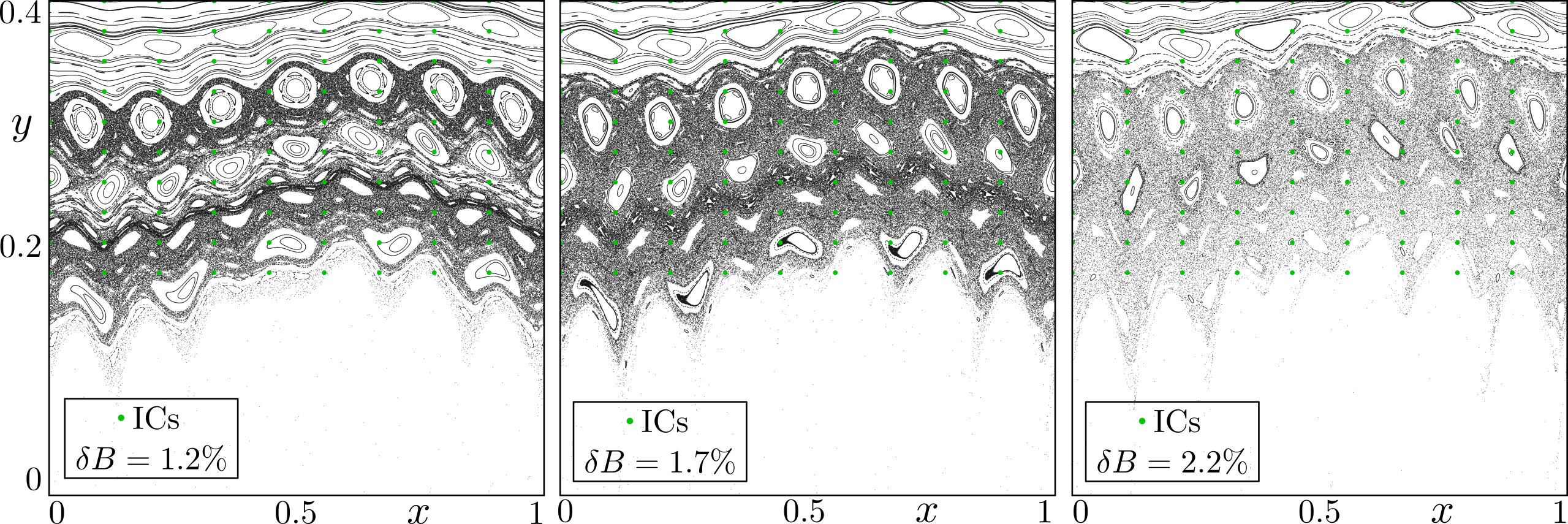}
\caption{Phase spaces of the UM considering $\delta B=1.2\%$ (left), $\delta B = 1.7\%$ (centre) and $\delta B = 2.2\%$ (right). The selected ICs are represented by the green points and their evolution was up to $2 \times 10^4$ iterations of the composed map $T^0_{\text{UM}} \circ T^1_{\text{UM}}$. All three phase spaces are drawn considering the region $x \in [0,1]$ and $y \in [0, 0.4]$.}
\label{fig:UM_3ps}
\end{figure}

Before analysing the $ER$ as a function of the control parameters, it is necessary to ensure that the escape conditions can be indeed satisfied by the evolved trajectories. Since only chaotic trajectories can escape, a suitable chaotic region must be found in the phase spaces. For that, Figs. \ref{fig:BM_3ps} and \ref{fig:UM_3ps} draw phase spaces for the BM and UM respectively, considering three different parameter values. These values correspond to both ends (higher and lower) of the defined ranges in Eq.\ \ref{eq:method_ranges} and a middle value. It is important to note how the phase space configuration changes while increasing the value of the control parameter.

Once the ranges of parameters of interest are established and the phase spaces are drawn, it is possible to identify a region where all given ICs can be placed within the chaotic sea. However, it is essential that this specific chaotic region is maintained throughout the entire parameter range\footnote{Identifying this robust chaotic region is a meticulous procedure specifically for the BM. As evidenced by Fig.\ \ref{fig:BM_3ps}, both the topology and the size of the phase space are highly sensitive to the parameter change.}. These conditions assured, only chaotic trajectories are evolved and, therefore, the escape conditions can be satisfied.

\begin{figure}[h!]
\centering
\includegraphics[scale=0.55]{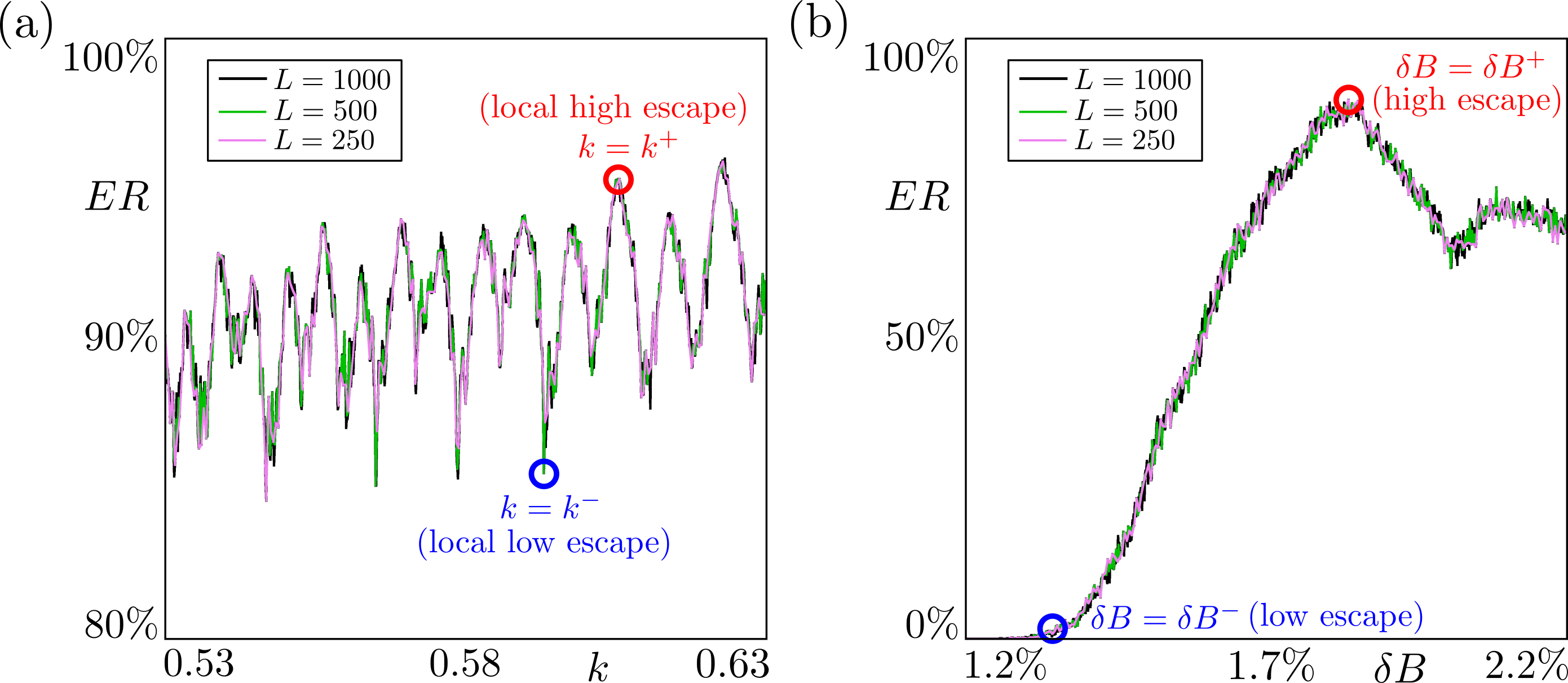}
\caption{Behaviour of the escape rate as a function of the control parameters for both the BM (a) and UM (b) models. Their defined ranges were divided into $L$ different parameters and, the colour lines show the behaviour of $1000$ (black), $500$ (green) and $250$ (violet) parameters.}
\label{fig:MTD_ER_parameter}
\end{figure}

First, for the BM, $M = 10^5$ ICs placed at $\xi_0 = 0$ and $\psi_0 \in [1-10^{-9},1-10^{-10}]$, were evolved up to $10^3$ iterations of the map. The computed $ER = ER(k)$, considering $L = 250$, $500$ and $1000$ parameters distributed at $k \in [0.53, 0.63]$, is shown in Fig.\ \ref{fig:MTD_ER_parameter}(a). It is worth noting that the general behaviour of $ER(k)$ is the same for the tree values of $L$.

For the UM, $M = 10^3$ ICs where placed at $x_0 \in [10^{-6},10^{-5}]$ and $y_0 = 0.3$ and evolved up to $10^5$ iterations. The computed $ER = ER(\delta B)$, considering $L = 250$, $500$ and $1000$ parameters distributed at $\delta B \in [1.2\%, 2.2\%]$, is shown in Fig.\ \ref{fig:MTD_ER_parameter}(b). It is also worth noting that the general behaviour of $ER(\delta B)$ is the same for the tree selected values of $L$.

In general terms, the escape analysis illustrated in Fig.\ \ref{fig:MTD_ER_parameter} yields precise values for the control parameters that indicate a configuration which enhances or restrains the escape. For the BM, $k = k^{-} = 0.5930$ and $k = k^{+} = 0.6056$ are the parameters of interest: $ER(k^{+})$ is the local high around $k\approx 0.6$, which not only provides better visualisation of the phase space but also is an adequate value as discussed in Sec.\ \ref{models:boozer} and, $ER(k^{-})$ the local low. For the UM, $\delta B = \delta B^{-} = 1.334\%$ and $\delta B = \delta B^{+} = 1.836\%$ are the parameters of interest: $ER(\delta B^{+})$ is the global high and, $ER(k^{-})$ is the global low that also indicates the first parameter value that escapes occurs.  

Furthermore, it is worth remarking that the general behaviour displayed in both panels of Fig.\ \ref{fig:MTD_ER_parameter} provides additional relevant interpretations. On one hand, the analysis of the BM shows a curious oscillatory behaviour that might be related to intrinsic dynamical structures, such as homoclinic tangles that arises and vanishes, around the saddle point. On the other hand, the analysis of the UM yields an expected growth while increasing the parameter value, however, there is a maximum followed by an arguably puzzling decay. Nevertheless, these interesting features would be further and properly investigated in other opportunities. 

Finally, once defined the values of interest $k^+$ and $k^-$ for the BM and, analogously $\delta B^{+}$ $\delta B^{-}$ for the UM, it is possible to thoroughly investigate their respective phase spaces via two of our original methods. Next section is devoted to presenting the results.

\section{Numerical results}
\label{results}
We discuss and present in this section our numerical results from the two selected methods to investigate the differences between the magnetic configurations of the models. In the first subsection, we employ the recurrence method outlined in \cite{Palmero2022} to pinpoint orbits that will experience stickiness on both models. Additionally, in the second subsection we investigate the average transient behaviour of escaping trajectories via the method outlined in \cite{deOliveira2022}.  

\subsection{Recurrence and stickiness}
Accordingly to \cite{Palmero2022}, by defining an ensemble of ICs, evolving them until a given maximum iteration time and computing the \emph{recurrence rate} of each orbit, it is possible to find particular trajectories that widely differ from the average behaviour. On this basis, it is possible to verify that orbits with high recurrence rates are the ones that experience \emph{stickiness}. Indeed, we verify the existence of stickiness separately for the BM and the UM as follows.

\subsubsection*{Single-null divertor map}
First, for the BM we selected an ensemble E formed by $M = 4.9 \times 10^7$ ICs uniformly distributed\footnote{A technical observation regarding the total number of ICs is in order: It is actually $M = m_x \times m_y$, with $m_x = m_y = 7 \times 10^3$ ICs; In that sense, the actual coordinates of the ICs are calculated over the intervals $\xi_0 \in [\xi_0^{\text{min}},\xi_0^{\text{max}}]/m_x$ and $\psi_0 \in [\psi_0^{\text{min}},\psi_0^{\text{max}}]/m_y$.} between $\xi_0^{\text{E}} \in [\xi^{u} - 10^{-10}, \xi^{u} + 10^{-10}]$ and $\psi_0^{\text{E}} \in [\psi^{u} - 10^{-10}, \psi^{u} + 10^{-10}]$, where $(\xi^u,\psi^u) = (0.00066855, 0.99774007)$ are the coordinates of the first Unstable Periodic Orbit (UPO) found within the chaotic portion of the phase space for $k = k^-$. 

Considering the phase space $k = k^+$, an analogous ensemble was placed again in the closest vicinity of the first UPO found within the chaotic separatrix; $\xi_0^{\text{E}} \in [\xi^{u} - 10^{-10}, \xi^{u} + 10^{-10}$ and $\psi_0^{\text{E}} \in [\psi^{u} - 10^{-10}, \psi^{u} + 10^{-10}]$, with $(\xi^u,\psi^u) = (0.00000001, 0.99698287)$. 

It is worth mentioning that, in practice, comparing the behaviour of trajectories from ensembles of ICs for different phase space configurations is a meticulous task. In that sense, both ensembles defined above were constructed considering the same number of ICs $M$, placed within small squares with the same width and height, centred around the computed coordinates (up to $10^{-8}$ numerical precision) of the first UPO found in their respective phase spaces. For $k = k^-$, it was found a period 29 UPO and, for $k = k^+$ a period 55 UPO. The UPOs' close vicinity is a convenient location because it warrants only chaotic orbits.

\begin{figure}[t!]
\centering
\includegraphics[scale=0.75]{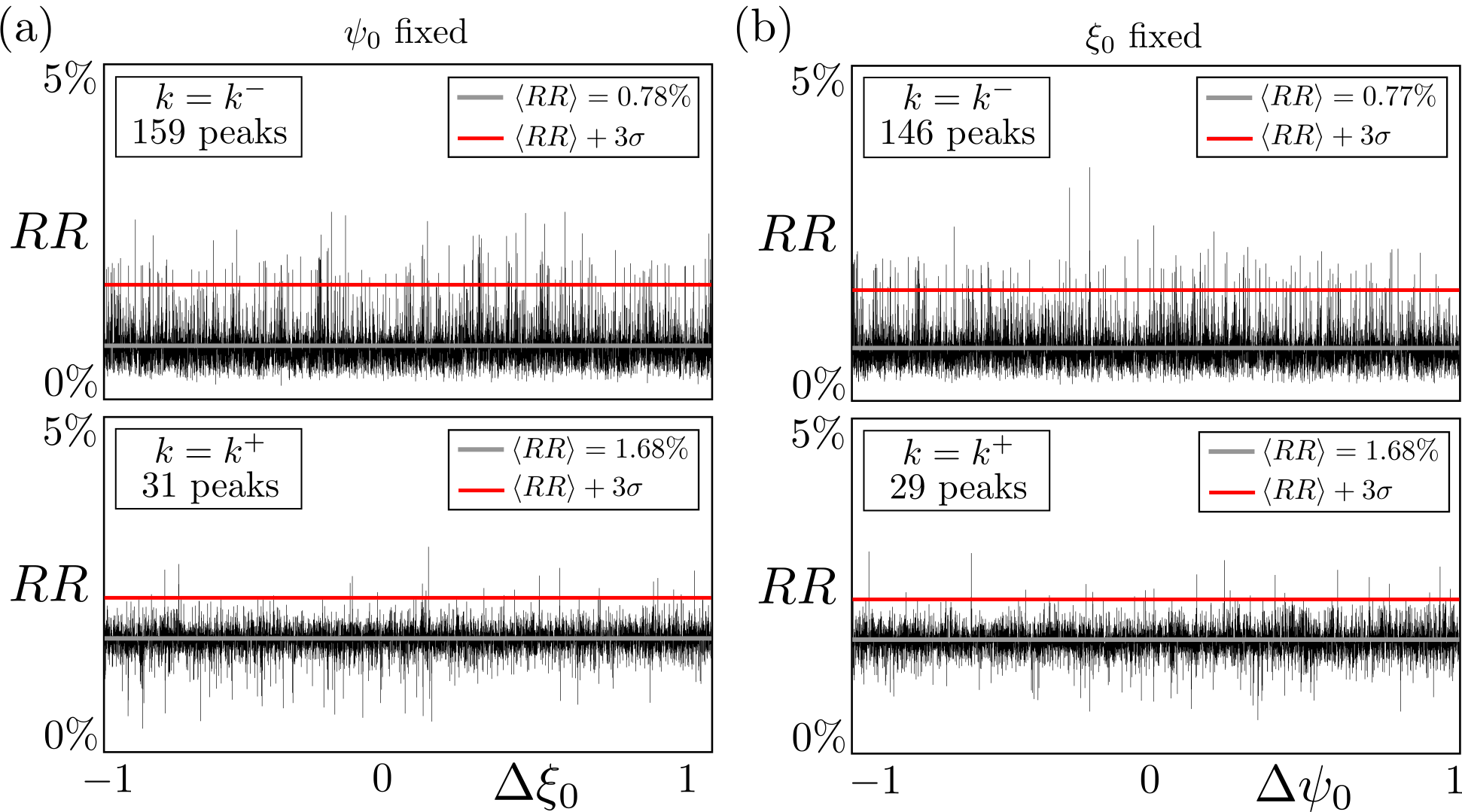}
\caption{$RR$ distributions as a function of the ICs for the BM. In (a) $\psi_0 = \psi^u$ is fixed, varying $\xi_0$ between the defined interval for both $k-$ (upper panel) and $k^+$ (lower panel). In (b) $\xi_0 = \xi^u$ is fixed, varying $\psi_0$ in the defined interval for $k-$ and $k^+$. Their corresponding values of $\langle RR \rangle$, the upper limit $\langle RR \rangle + 3\sigma$ depicted by the red line and the count of peaks are displayed in all panels.}
\label{fig:REC_bm_rr_ics}
\end{figure}

Once the ensembles are well-defined, we begin the recurrence analysis considering all $M = 4.9 \times 10^7$ ICs, evolved until $N = 10^4$ iterations and the recurrence threshold distance $\varepsilon = 0.01$. Figure \ref{fig:REC_bm_rr_ics} presents the first results of the computed $RR$ for all trajectories in the ensembles.

Since the defined ICs were placed on intervals in both axis, Fig.\ \ref{fig:REC_bm_rr_ics}(a) and (b) separates the analysis fixing $\psi_0$ and $\xi_0$ respectively and varying the other coordinate within the ensemble's size. We define relative distances $\Delta\xi_0$ and $\Delta \psi_0$, taking as a reference the coordinates of their corresponding UPOs $(\xi^u,\psi^u)$, as $\Delta (\xi_0,\psi_0) = [(\xi_0^\text{E}, \psi_0^\text{E}) - (\xi^u,\psi^u)] \times 10^{10}$. Then, the upper and lower panels of Fig.\ \ref{fig:REC_bm_rr_ics} (a) show the recurrence rate distribution over the interval of ICs on $\xi$ for $k^-$ and $k^+$ respectively. Analogously the panels of Fig.\ \ref{fig:REC_bm_rr_ics} (b) display the same distributions now over the interval of ICs on $\psi$.

By defining an upper limit, calculated by $\langle RR \rangle + 3\sigma$ i.\ e.\ the average recurrence rate over the ensemble of ICs plus three times the standard deviation of the average, it is possible to perform a 3$\sigma$ detection on how many peaks are in the distributions. A peak is associated with a specific value of IC that provides a highly recurrent trajectory.

The foremost important outcome of Fig.\ \ref{fig:REC_bm_rr_ics} is the difference between the number of peaks for $k^-$ and $k^+$. It was established that $k^-$ is the value of the perturbation strength that gives a phase space configuration which restrains escape orbits. In that regard, it is reasonable to expect that this phase space configuration would present many trajectories that experience stickiness and, due to this trapping time, they might not escape the system until the considered evolution time. The opposite is to be expected for $k^+$, being associated with a phase space that enhances escaping trajectories. All four distributions shown in Fig.\ \ref{fig:REC_bm_rr_ics} and their respective peak counts notably agree with these expectations. 

\begin{figure}[h!]
\centering
\includegraphics[scale=0.7]{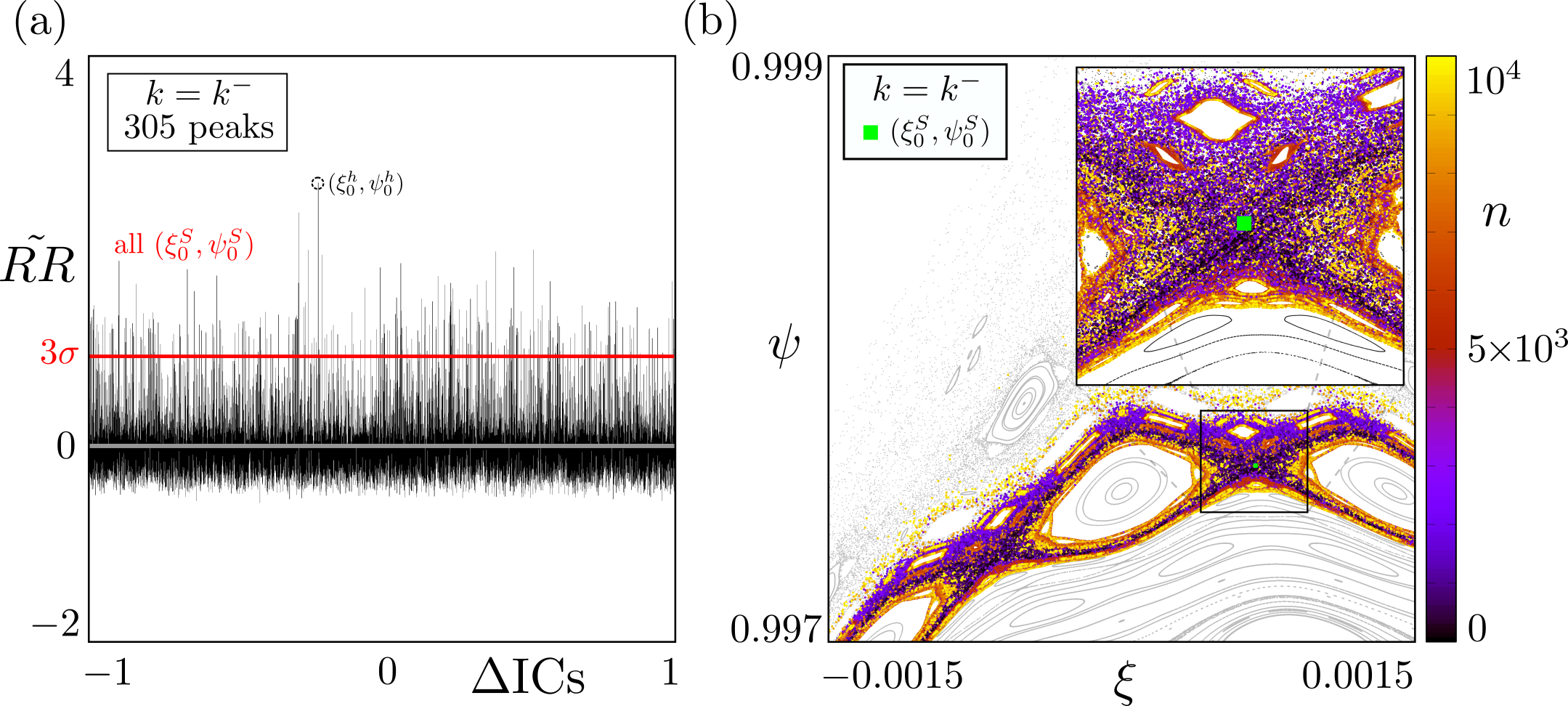}
\caption{(a) $\tilde{RR}$ distribution as a function of all $M=4.9\times10^7$ ICs evolved until $N=10^4$ iterations of the BM, considering $k=k-$ and the threshold distance $\varepsilon = 0.01$. The IC $(\xi_0^h,\psi_0^h)$ with the highest $RR$ is marked by the dashed circle; (b) 305 trajectories evolved from all $(\xi_0^S,\psi_0^S)$ determined from the 3$\sigma$ limit in (a). The position of the ensemble is depicted by the green square (out of scale) and the inset shows the amplified region inside the black square.} 
\label{fig:REC_bm_rr_ps_k-}
\end{figure}

Furthermore, Fig.\ \ref{fig:REC_bm_rr_ps_k-} (a) combines prior recurrence analyses for $k^-$ on both intervals into one considering $\tilde{RR}$ as a function of $\Delta IC$, where we define $\tilde{RR} = RR - \langle RR \rangle$ as the corrected $RR$ in respect to its ensemble average, and $\Delta IC = \Delta\xi_0 + \Delta \psi_0$ as the combined intervals for the ICs. The upper limit is depicted by the red line at $3\sigma$ since the $\tilde{RR}$ distribution is null at $\langle RR \rangle$ by construction. All ICs that, when evolved up to $N$ iterations, produce trajectories with computed $RR$ larger than the upper limit are suitable candidates to present stickiness, so we label these ICs as $(\xi_0^S,\psi_0^S)$.  

Essentially, the analysis of the $RR$ distribution over the ICs provides a subset $S$ of the ensemble E, composed solely of ICs that, when evolved, will produce highly recurrent trajectories. For the case of $k = k^-$ the subset $S$ is formed by 305 ICs and their evolution is shown in Fig.\ \ref{fig:REC_bm_rr_ps_k-} (b). These 305 trajectories do not visit the upper regions of the chaotic separatrix, being dynamically trapped in the chaotic area between the period 29 chain of islands and yet, finishing their evolution by visiting the neighbourhood of all embedded periodic islands. By this numerical observation, these are indeed trajectories that experience stickiness phenomena.  

\begin{figure}[h!]
\centering
\includegraphics[scale=0.7]{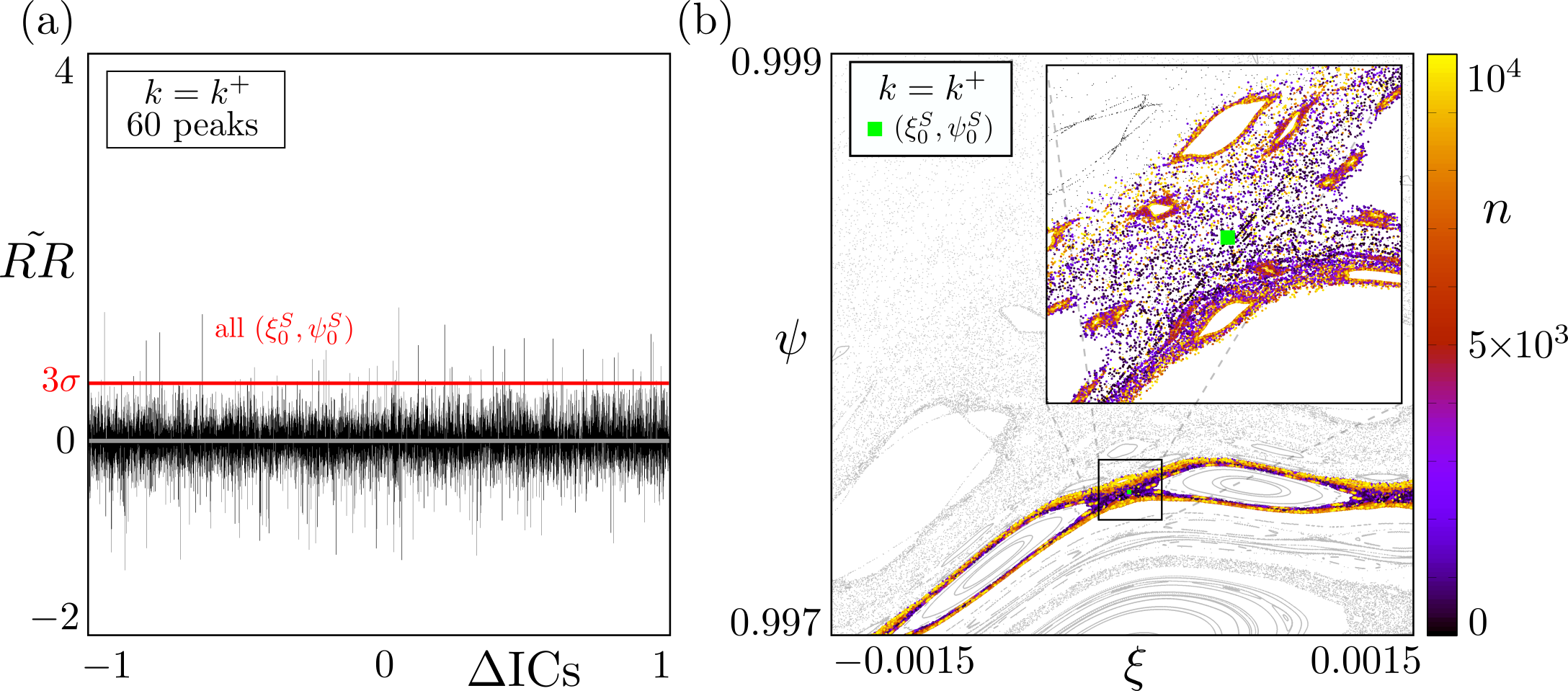}
\caption{(a) $\tilde{RR}$ distribution as a function of all $M=4.9\times10^7$ ICs evolved until $N=10^4$ iterations of the BM, considering $k=k+$ and the threshold distance $\varepsilon = 0.01$. (b) 60 trajectories evolved from all $(\xi_0^S,\psi_0^S)$ determined in (a). The position of the ensemble is depicted by the green square (out of scale) and the inset shows the amplified region inside the black square.} 
\label{fig:REC_bm_rr_ps_k+}
\end{figure}

\begin{figure}[b!]
\centering
\includegraphics[scale=0.70]{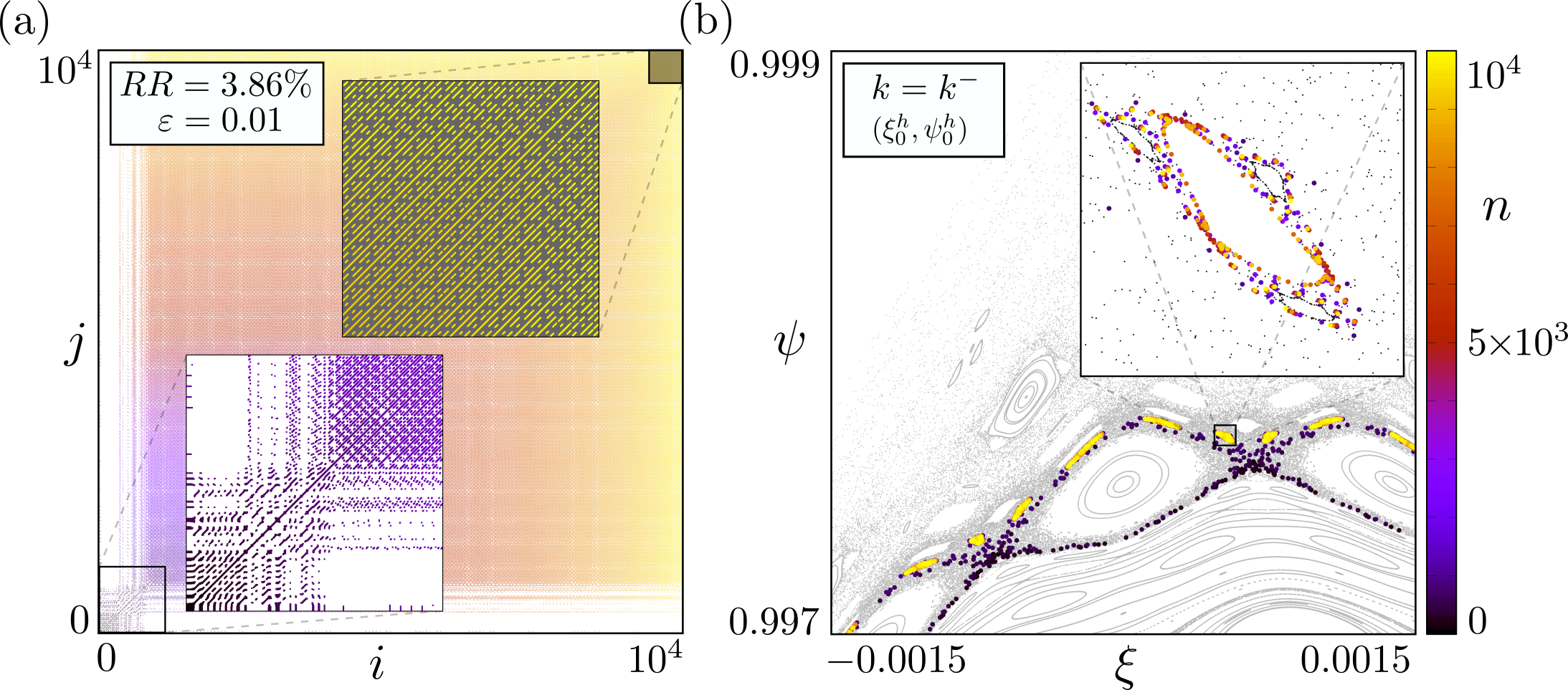}
\caption{(a) RP with $\varepsilon = 0.01$ for the trajectory started from $(\xi_0^h,\psi_0^h)$. The insets show amplifications of their respective regions in the black squares. Both axes are set in the same colour gradient of the trajectory's evolution; (b) Phase space for $k = k^-$ (grey on the background) along with the same orbit. The inset shows the region around the small periodic island.} 
\label{fig:REC_bm_highest_rr}
\end{figure}

In the same direction, Fig.\ \ref{fig:REC_bm_rr_ps_k+} (a) presents the results of the distribution $\tilde{RR} \times \Delta IC$ considering $k = k^+$, where the subset $S$ is roughly three times smaller than the previous analysis, composed of 60 ICs. Nevertheless, the evolution of these special trajectories is shown in Fig.\ \ref{fig:REC_bm_rr_ps_k+} (b).  

The evolution of the subset $S$ displayed in Fig.\ \ref{fig:REC_bm_rr_ps_k+} (b) displays all trajectories in the initial chaotic surroundings, never visiting upper regions of the phase space. The inset reveals that these 60 orbits are trapped around many small periodic structures, indicating stickiness once more.

The final numerical observation for the BM is shown in Fig.\ \ref{fig:REC_bm_highest_rr}, where we evolve the IC $(\xi_0^h,\psi_0^h)$ with the highest $RR$ from the distribution in Fig.\ \ref{fig:REC_bm_rr_ps_k-} (a) and we display its RP considering the same $N = 10^4$ and $\varepsilon = 0.01$ established throughout the analysis. The strong quasi-periodic is evident in the RP displayed in Fig.\ \ref{fig:REC_bm_highest_rr} (a). The first inset shows the initial one-thousand iterations, where a chaotic transient-like evolution can be clearly observed before the strong stickiness region. The last inset highlights the periodic patterns within the final five-hundred iterations. While compared to Fig.\ \ref{fig:REC_bm_highest_rr} (b), we observe the exact place in which the orbit gets trapped. The inset shows more clearly this fine chaotic vicinity around the periodic island. 

\subsubsection*{Ergodic magnetic limiter map}

\begin{figure}[h!]
\centering
\includegraphics[scale=0.75]{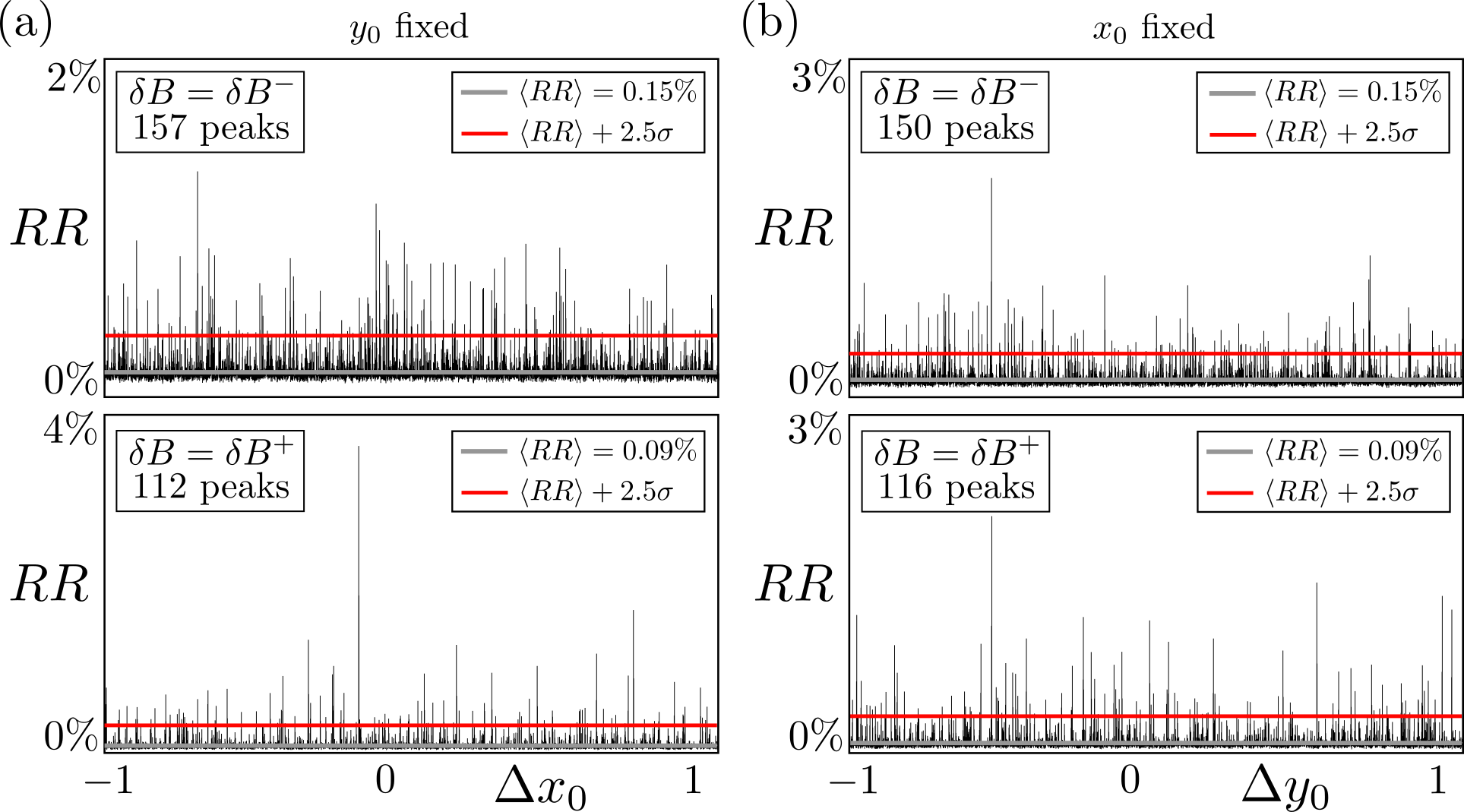}
\caption{$RR$ distributions as a function of the ICs for the UM. In (a) $y_0 = y^u$ is fixed, varying $x_0$ between the defined interval for both $\delta B^{-}$ (upper panel) and $\delta B^+$ (lower panel). In (b) $x_0 = x^u$ is fixed, varying $y_0$ in the defined interval for $\delta B^{-}$ and $\delta B^+$. Their corresponding values of $\langle RR \rangle$, the upper limit $\langle RR \rangle + 2.5\sigma$ depicted by the red line and the count of peaks are displayed in all panels.}
\label{fig:REC_um_rr_ics}
\end{figure}

As the recurrence-based detection approach proved to be effective in identifying stickiness for the BM, we employ a similar analysis for the UM. An ensemble E formed by $M = 2.5 \times 10^7$ ICs is uniformly distributed between small intervals in both $x$ and $y$-axis; $x_0^{\text{E}} \in [x^{u} - 10^{-10}, x^{u} + 10^{-10}]$ and $y_0^{\text{E}} \in [y^{u} - 10^{-10}, y^{u} + 10^{-10}]$, where $(x^u,y^u)$ are the coordinates of the period 7 UPO found in the upper chaotic area of the phase space. Unlikely the BM, the phase space of the UM is more robust to changes in the control parameter of the map, given by the relative perturbation of the magnetic field $\delta B$, making it easier to find a suitable location for positioning E while considering the two different configurations $\delta B^-$ and $\delta B^+$. Specifically, E is centred at $(x^u,y^u) = (0.420363188, 0.321734626)$ for $\delta B^-$ and at $(x^u,y^u) = (0.420594827, 0.321787496)$ for $\delta B^+$. 

Now that the ensembles are well-defined we analyse all $M = 2.5 \times 10^7$ ICs, evolved until $N = 10^4$ iterations, considering the recurrence threshold distance $\varepsilon = 0.005$ to compute their corresponding $RR$\footnote{To further investigate the adaptability of our method, we consider now a different value of the recurrence threshold distance $\varepsilon = 0.005$ and we perform a 2.5$\sigma$ detection to pinpoint special ICs for stickiness in the UM.}. We present in Fig.\ \ref{fig:REC_bm_rr_ics} the results for the $RR$ distribution over $x_0$ and $y_0$ intervals separately and for the two configurations $\delta B^-$ and $\delta B^+$ of the UM. 

Figure \ref{fig:REC_um_rr_ics} shows that, although the number of peaks for $\delta B^-$ is higher than for $\delta B^+$ in both $x$ and $y$ distributions, the difference is not as high as for $k^-$ and $k^+$ previously shown in Fig. \ref{fig:REC_bm_rr_ics} for the BM. Nevertheless, the relatively high number of peaks for the configuration $\delta B^+$ indicates that we will find strong stickiness in the corresponding phase space as well.

\begin{figure}[h!]
\centering
\includegraphics[scale=0.70]{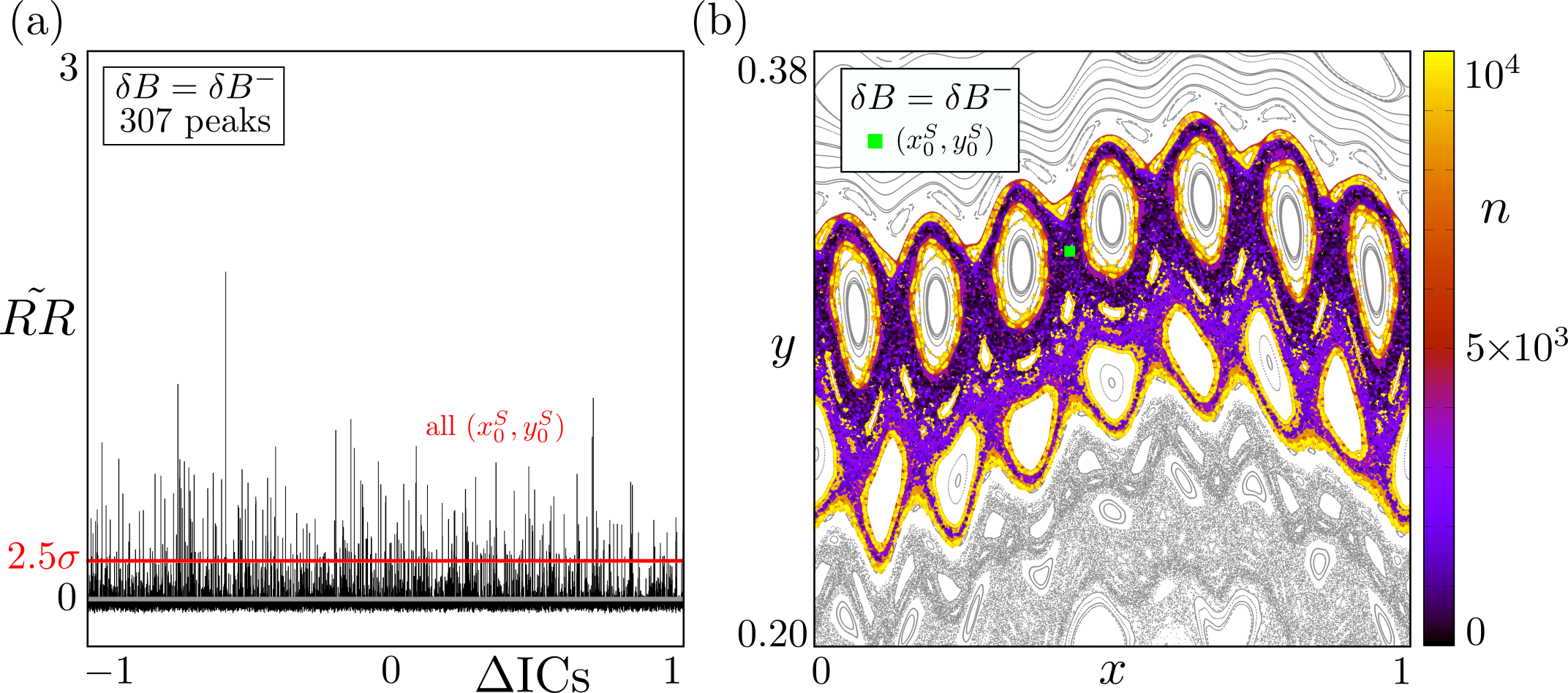}
\caption{(a) $\tilde{RR}$ distribution as a function of all ICs evolved until $N=10^4$ iterations of the UM, considering $\delta B^-$ and the threshold distance $\varepsilon = 0.005$; (b) 307 trajectories evolved from all $(x_0^S,y^S)$ determined from the 2.5$\sigma$ detection in (a). The position of the ensemble is depicted by the green square (out of scale).} 
\label{fig:REC_um_rr_ps_b-}
\end{figure}

From both upper panels of \ref{fig:REC_um_rr_ics} we found 307 ICs that compose a special subset $S$ of our initial ensemble E. These special ICs are labelled $(x_0^S,y_0^S)$ in Fig.\ \ref{fig:REC_um_rr_ps_b-} (a), where we combine previous distributions on both intervals into one while considering the aforementioned corrected $\tilde{RR}$ recurrence rate. In Fig.\ \ref{fig:REC_um_rr_ps_b-} (b) we display the evolution of $(x_0^S,y_0^S)$ in the corresponding phase space configuration $\delta B^-$, where the colour gradient from the number of iterations reveals that most of these special trajectories spend an expressive amount of time around the KAM islands embedded in this upper region of the phase space.

\begin{figure}[h!]
\centering
\includegraphics[scale=0.70]{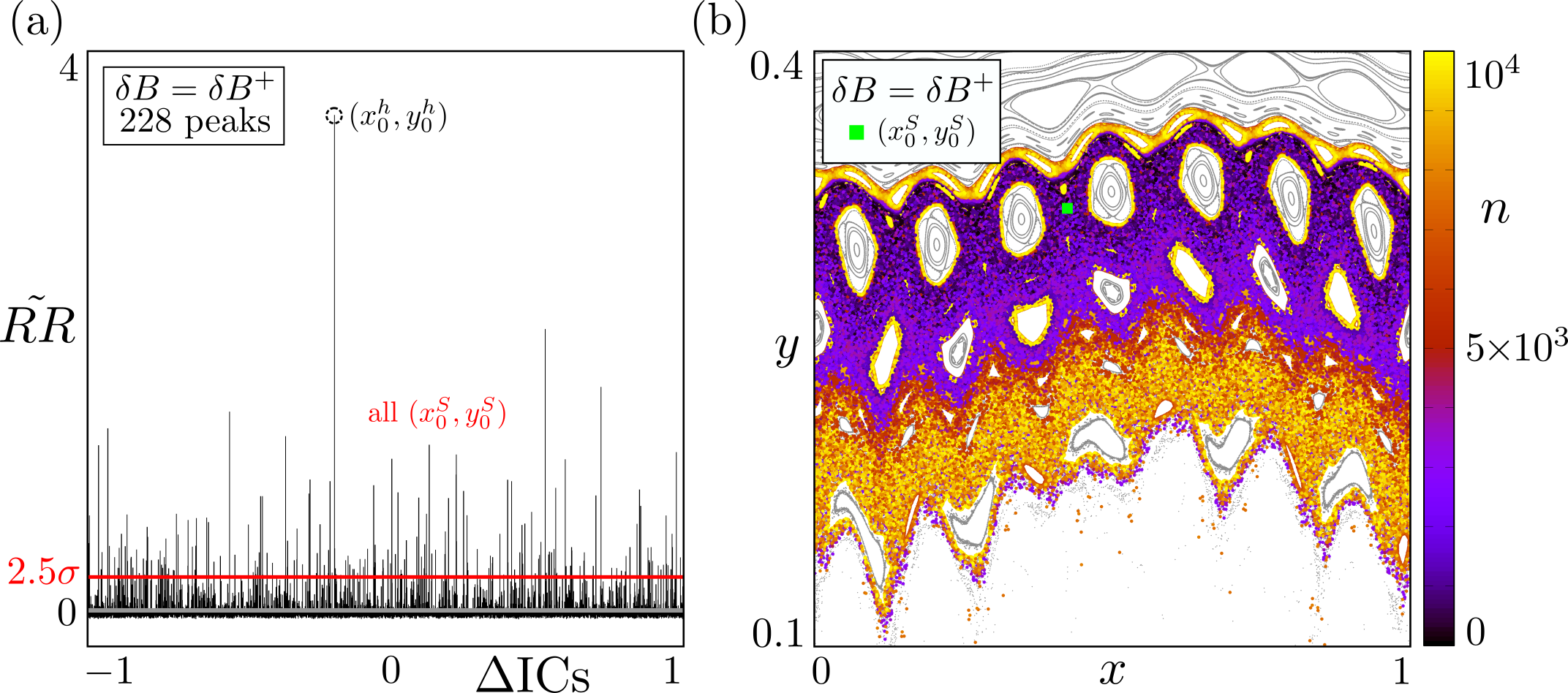}
\caption{(a) $\tilde{RR}$ distribution as a function of all ICs evolved until $N=10^4$ iterations of the UM, considering $\delta B^+$ and the threshold distance $\varepsilon = 0.005$. The IC $(x_0^h,y_0^h)$ with the highest $RR$ is marked by the dashed circle; (b) 228 trajectories evolved from all $(x_0^S,y^S)$ from the 2.5$\sigma$ detection in (a). The position of the ensemble is depicted by the green square (out of scale).} 
\label{fig:REC_um_rr_ps_b+}
\end{figure}

\begin{figure}[h!]
\centering
\includegraphics[scale=0.70]{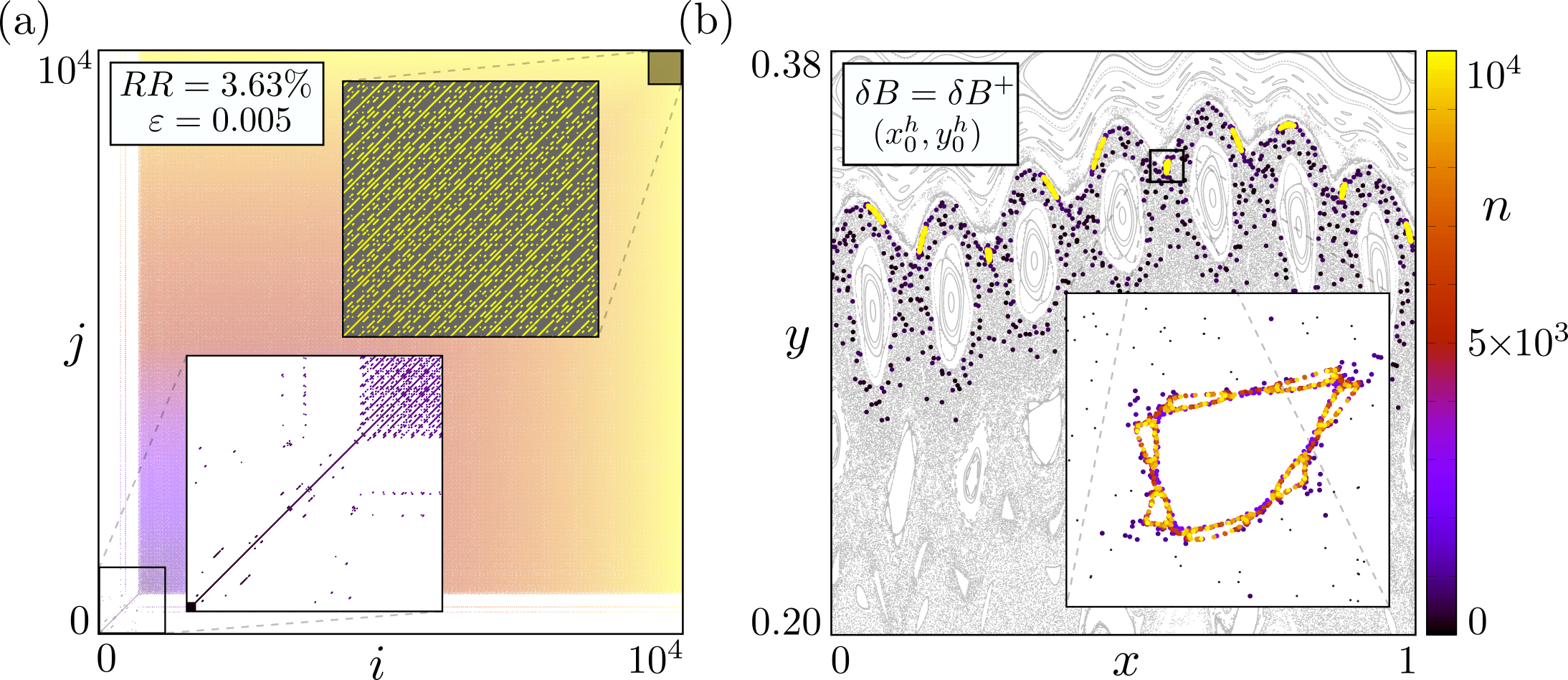}
\caption{(a) RP with $\varepsilon = 0.005$ for the trajectory started from $(x_0^h,y_0^h)$. The insets show amplifications of their respective regions in the black squares. Both axes are set in the same colour gradient of the trajectory's evolution; (b) Phase space for $\delta B = \delta B^+$ (grey on the background) along with the same orbit. The inset shows the region around the small periodic island.} 
\label{fig:REC_um_highest_rr}
\end{figure}

Moreover, Fig.\ \ref{fig:REC_um_rr_ps_b+} shows the same results now considering the configuration $\delta B^+$. The 2.5$\sigma$ detection in (a) selects 228 ICs, now labelled $(x_0^S,y^S)$, that are evolved up to $10^4$ iterations in panel (b). Additionally, Fig.\ \ref{fig:REC_um_rr_ps_b+} (a) highlights a special IC $(x_0^h, y_0^h)$ that presents a computed $RR \approx 3.6\%$ about forty times the average over the ensemble $\langle RR \rangle = 0.09\%$, meaning that an orbit started from $(x_0^h, y_0^h)$ is an extremely recurrent chaotic trajectory that will be further inspected in Fig.\ \ref{fig:REC_um_highest_rr}.

The phase space with the sub-ensemble's $S$ evolution depicted in Fig.\ \ref{fig:REC_um_rr_ps_b+} (b) is again marked by trajectories that spend much time around periodic structures, all highlighted by the yellow points, indicating stickiness phenomena. Furthermore, it is interesting to note a pronounced separation between the upper region, mostly in purple-like colours, and the lower region in red/yellow. This separation is often associated with strong transport barriers found in phase spaces of general non-linear symplectic maps \cite{Viana2021}, which once analysed from the magnetic confinement point of view, have important implications for tokamaks and plasma-wall interactions \cite{Samm2010}.   

Finally, Fig.\ \ref{fig:REC_um_highest_rr} presents the final numerical observation from the recurrence-based detection approach for the UM. Analogous to what was presented for the BM in Fig.\ \ref{fig:REC_bm_highest_rr}, the RP related to the special IC $(x_0^S,y^S)$ is shown in panel (a), along with its evolution on the phase space for $\delta B^+$ in panel (b). 

The RP in Fig.\ \ref{fig:REC_um_highest_rr} (a) was constructed considering the same recurrence distance threshold $\varepsilon = 0.005$ and the maximum number of iterations $N = 10^4$ established in all analyses for the UM. The strong quasi-periodic is once more evident, along with the chaotic transient-like region observed in the first thousand iterations, highlighted in the initial inset. After this transient time, the stickiness region dominates the entire RP and, in the final inset, we depict the periodic patterns for the last five-hundred iterations. In Fig.\ \ref{fig:REC_bm_highest_rr} (b) we observe the behaviour of the same trajectory on the phase space, where the trapping region is highlighted in yellow. The inset shows more clearly the narrow chaotic vicinity around the island.

\subsection{Transient motion}
The transient motion analysis offers a visual aid for the hidden transient dynamics of escaping trajectories. The method is detailed described in \cite{deOliveira2022}. Essentially, a single chaotic orbit will visit only a small portion of the available chaotic area before escaping, making it hard to understand the behaviour of escape orbits in the phase space by looking solely at the \emph{transient measure} $\nu$. We define, thereby, the \emph{mean transient measure} $\langle \nu_i \rangle$, as the average of the transient measure on an ensemble composed by $M$ ICs. Then, we employ the method on numerical simulations of both BM and UM models. The results are shown separately below.

\subsubsection*{Single-null divertor map}
The topology of the magnetic field lines induced by the single-null poloidal divertor, represented via the characteristic phase space of the BM shown in Fig.\ \ref{fig:BM_phase_space}, presents a saddle point at $(\xi^{\star}, \psi^{\star}) = (0,1)$. Accordingly, the position of the divertor target is considered $\psi_{\text{target}} = 1.0$ to be the nearest to the saddle point, imposing the escape condition $\psi_n < 1.0 < \psi_{n+1}$ to the map equations. 

\begin{figure}[h!]
\centering
\includegraphics[scale=0.75]{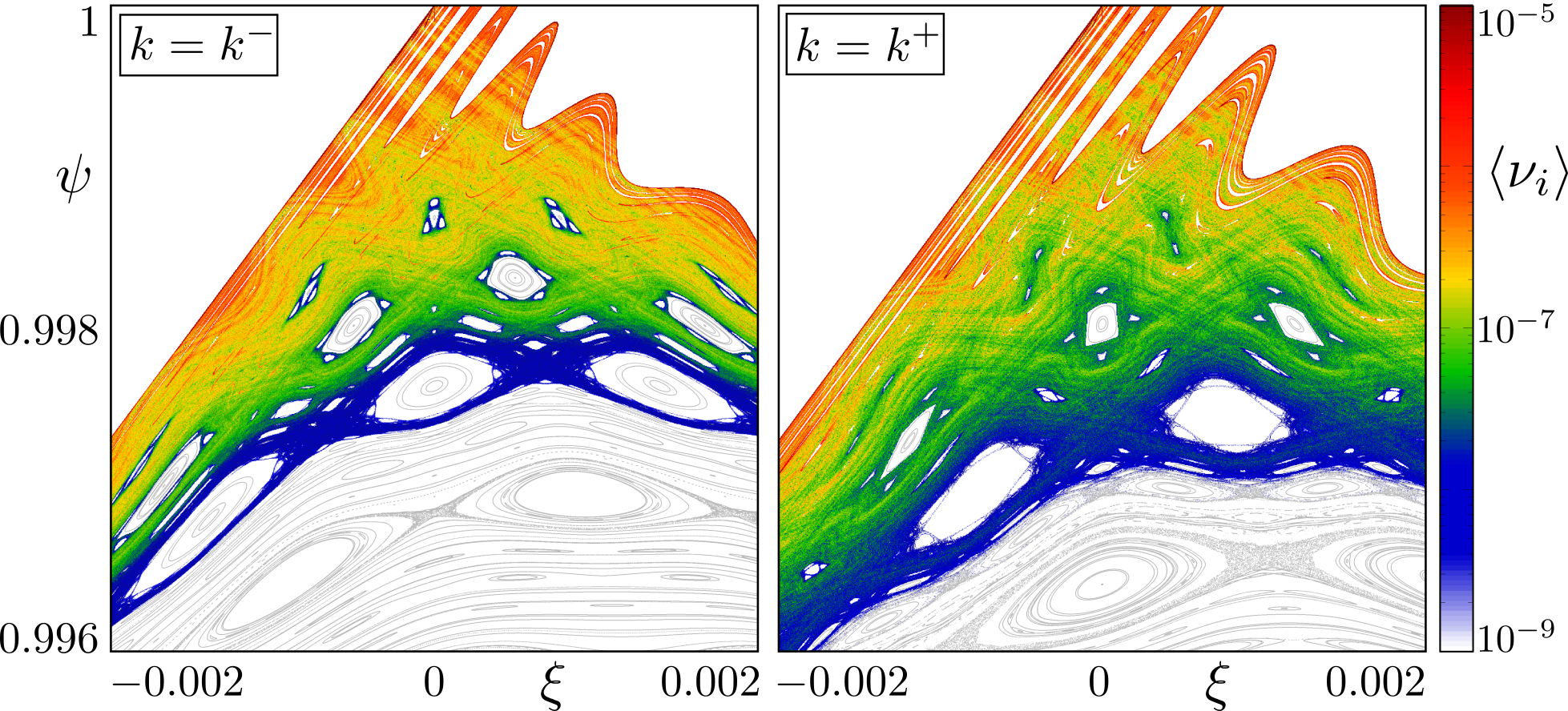}
\caption{Profiles of the mean transient measure $\langle \nu_i \rangle$ in logarithmic scale for the BM, calculated on a $1024\times1024$ grid, considering $k=k^{-}=0.5930$ (left) and $k=k^{+}=0.6056$ (right). Their respective phase spaces are depicted in grey on the background. The ensemble of ICs and the colour range for the mean transient measure were kept the same for both cases.}
\label{fig:TMA_bm_profiles}
\end{figure}

Once defined the escape condition, we select an ensemble of $M = 10^5$ ICs, uniformly distributed in a dense small line positioned at $\xi_0 = 0$, $\psi_0 \in [1 - 1\times10^{-4}, 1 - 1\times10^{-6}]$ evolved up to $10^7$ iterations of the map. For the computation of the transient measure, it was considered $\varepsilon = 1024$ as the side-length of the boxes that compose a grid over the region of interest $\xi \in [-0.0025, 0.0025]$ and $\psi \in [0.995,1.0]$. Then, we compute the profile of the mean transient measure $\langle \nu_i \rangle$ throughout the phase space of the BM, considering the two special values of the control parameter: $k = k^{-}$, the perturbation strength that gives a low escape rate and; $k = k^{+}$, the perturbation strength that gives a high escape rate. Figure \ref{fig:TMA_bm_profiles} displays the results.  

From Fig.\ \ref{fig:TMA_bm_profiles} we readily observe that the $\langle \nu_i \rangle$ profile, depicted by the logarithmic colour scale, highlights the differences between both cases. As expected, since $k^{+} > k^{-}$, the available chaotic portion of the phase space for $k=k^{+}$ is larger than the phase space for $k=k^{-}$. However, the gradient from the colour scales provides better insights into how the chaotic orbits are experiencing these chaotic regions before escaping the system. On one hand, for $k^{-}$ escape orbits frequently visits regions around upper smaller islands, as depicted by the surrounding blue colours, strongly indicating the presence of stickiness. On the other, $k^{+}$ presents a relatively thinner profile, where most of the previous islands are already destroyed and the gradient from yellow to dark red reveals more visible gaps (in white) farther from the saddle. 

\begin{figure}[h!]
\centering
\includegraphics[scale=1.0]{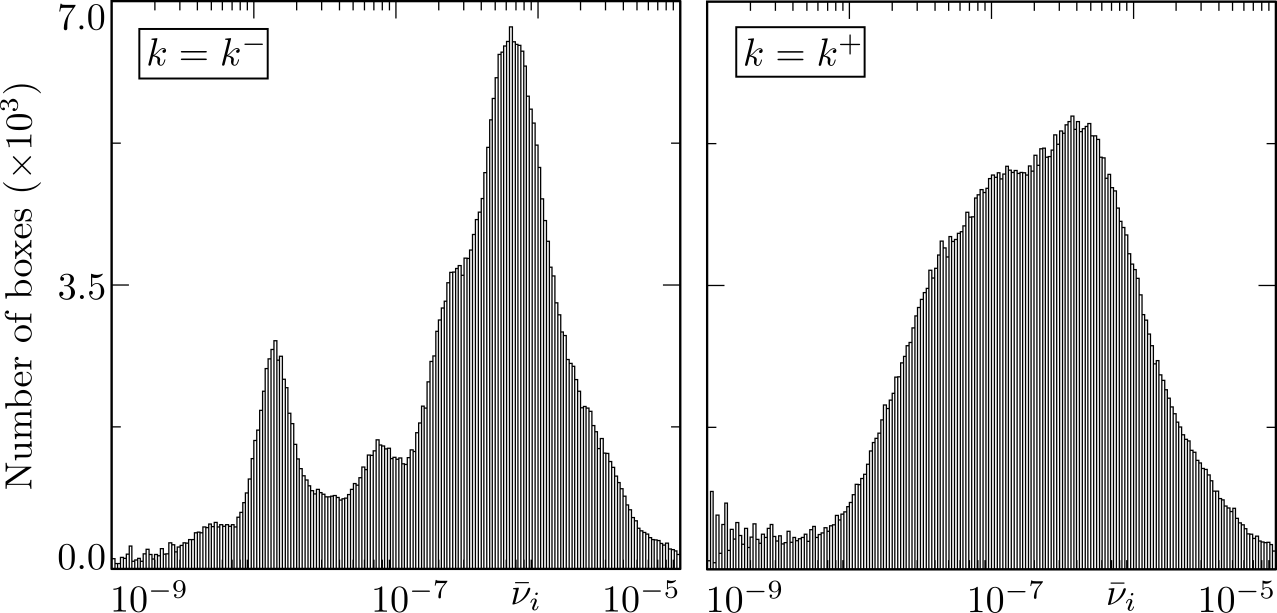}
\caption{Histogram distributions of the mean transient measure $\langle \nu_i \rangle$ for the escape trajectories of the BM, considering both $k=k^{-}=0.5930$ (left) and $k=k^{+}=0.6056$ (right). All parameters were kept the same as for Fig.\ \ref{fig:TMA_bm_profiles}.}
\label{fig:TMA_bm_histograms}
\end{figure}

Before diving into the comparison between the complicated structures uncovered by the profile and the correspondingly invariant manifolds, it is possible to statistically investigate the mean transient measure profiles by computing their histograms. Considering only the boxes visited at least once by the simulated dynamics for both cases, the result is shown in Fig.\ \ref{fig:TMA_bm_histograms}.  

The calculated histogram distributions stress the different transient behaviours that emerge from the complex dynamical scenario of the system, especially while comparing different perturbation strengths. Confronting both panels of Fig.\ \ref{fig:TMA_bm_histograms} to their respective profiles on the phase space, shown in Fig.\ \ref{fig:TMA_bm_profiles}, we note that, indeed, orbits within $k=k^{-}$ frequently visit regions that are not available in $k=k^{+}$. The smaller peak around $\langle \nu_i \rangle \approx 10^{-8}$ indicates a relatively high concentration for all possible paths in areas coloured blue (respective colour for $\langle \nu_i \rangle \approx 10^{-8}$) in Fig.\ \ref{fig:TMA_bm_profiles}. Still in $k=k^{-}$, the higher peak, around $\langle \nu_i \rangle \approx 10^{-6}$ is sharper compared to the broader distribution for $k=k^{+}$. This suggests a somewhat contra-intuitive realisation that although the phase space of $k^{-}$ presents a smaller chaotic region compared to $k^{+}$, escaping chaotic trajectories from $k^{-}$ are experiencing thoroughly most of the available regions and, in that sense, enabling stickiness phenomena, in comparison to a more erratic visitation of the orbits in $k^{+}$. 

\begin{figure}[h!]
\centering
\includegraphics[scale=1.0]{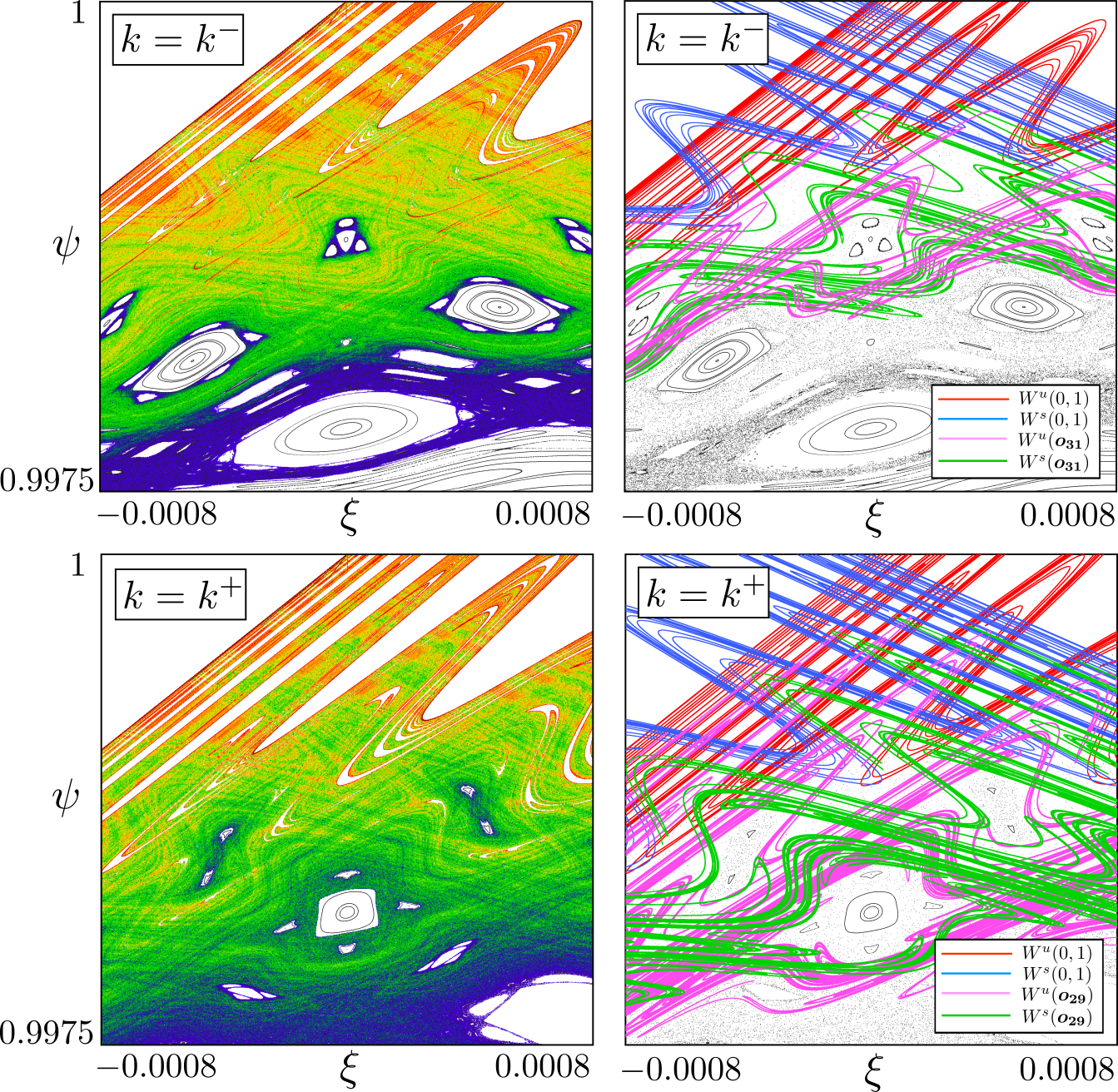}
\caption{Comparison between the $\langle \nu_i \rangle$ profile (left panels) and invariant manifolds $W^{s,u}$ (right panels) associated with the saddle point and nearest UPOs of interest. Their respective phase spaces are depicted in grey on the background. All parameters and the colour scale (omitted) are the same as in Fig.\ \ref{fig:TMA_bm_profiles}.}
\label{fig:TMA_bm_mani}
\end{figure}

Finally, as a last visual investigation for the BM, we compare the uncovered structural details shown in Fig.\ \ref{fig:TMA_bm_profiles} to the relevant invariant manifolds present in both phase spaces. As detailed in \cite{deOliveira2022}, we use the method proposed by Ciro {\it et al} \cite{Ciro2018} to calculate and trace the selected manifolds. To further improve the visualisation, we consider an amplified region $\xi \in [-0.0008, 0.0008]$ and $\psi \in [0.9975,1.0]$ where the respective details are more evident. Figure \ref{fig:TMA_bm_mani} shows, on the left panels, the computed profile throughout this region of the phase space and, on the right, the same region covered by stable and unstable invariant manifolds associated with the saddle $(\xi^{\star}, \psi^{\star}) = (0,1)$ and the nearest UPOs from the last visible chain of islands. For $k^{-}$, the nearest UPO is associated with a period 31 chain of islands and, for $k^{+}$ it is a period 29 UPO.

The colour gradient of $\langle \nu_i \rangle$ depicts, in both cases, complex geometrical structures, formed by seemingly erratic curves, that are embedded in their respective phase spaces. These structures accurately agree with the invariant manifolds shown in the right panels of Fig.\ \ref{fig:TMA_bm_mani}. Curiously, the structures revealed by the mean transient measure are consistent not only with the unstable manifolds but also with the stable ones. Moreover, while comparing both cases of $k$, it is possible to note that the traced manifolds for $k^+$ seem more interconnected\footnote{As a technical remark: It is known that, since these invariant manifolds are composed by infinite sets, both homo/heteroclinic intersections are, by theoretical definition, also invariant and infinite sets. That being said, we discuss the interconnectivity of these manifolds focusing only on what was possible to compute in our numerical simulations and display in the figures.} than the ones presented in $k^-$. Of course, knowing that the manifolds are often related to transport channels, the strong interconnected structures behind the phase space for $k^{+}$ might explain why it is a configuration that enhances the escape, differently from what is observed for $k^{-}$.

\subsubsection*{Ergodic magnetic limiter map}

\begin{figure}[h!]
\centering
\includegraphics[scale=0.75]{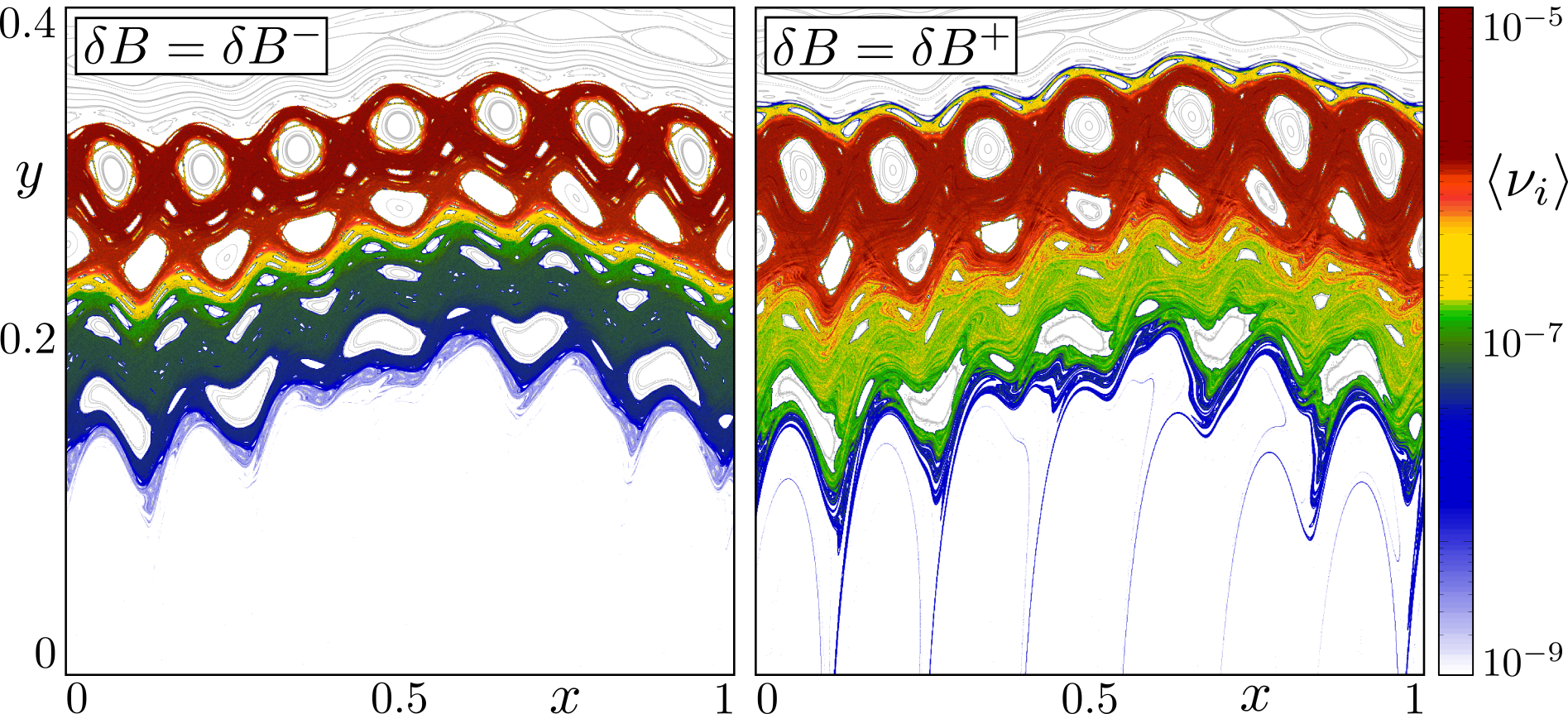}
\caption{Profiles of the mean transient measure $\langle \nu_i \rangle$ in logarithmic scale for the UM, calculated on a $1024\times1024$ grid, considering $\delta B=\delta B^{-}=1.334\%$ (left) and $\delta B=\delta B^{+}=1.836\%$ (right). Their respective phase spaces are depicted in grey on the background. The ensemble of ICs and the colour range for the mean transient measure were kept the same for both cases.}
\label{fig:TMA_um_profiles}
\end{figure}

Analogously to what was presented for the BM, depending on the values for the perturbation strength, provided either by $\delta B$ or $\delta I$, it was shown that escape field lines may be found, considering the escape condition $y_n < 0 < y_{n+1}$, meaning that a field line crossed the inner wall at $y = 0$. 

To numerically calculate the transient measure and the respective mean transient profile, we select an ensemble of $M = 10^4$ ICs, uniformly distributed in a dense small square in $x_0 \in [-1 \times 10^{-7}, 1 \times 10^{-7}]$ and $y_0 \in [0.3 - 1 \times10^{-7}, 0.3 + 1\times10^{-7}]$ evolved up to $10^5$ iterations. Only the lower region $y < 0.4$ with $x \in [0,1]$ was selected for the analysis, considering $\varepsilon = 1024$ as the boxes' side-length of the grid in this region. Then, the profile of the mean transient measure $\langle \nu_i \rangle$ for the UM was computed for the two special values: $\delta B = \delta B^{-}$, the perturbation strength that gives a low escape rate and; $\delta B = \delta B^{+}$, the perturbation strength that gives a high escape rate. Figure \ref{fig:TMA_um_profiles} displays the results.  

It is possible to readily observe that the $\langle \nu_i \rangle$ profile, depicted by the logarithmic colour scale in Fig.\ \ref{fig:TMA_um_profiles}, highlights the differences between both cases similarly to results for the BM. However, one main difference between the simulation of the UM and the BM is the selected location for the ensemble of ICs. For the UM, ICs are located at $y \approx 0.3$, far from the escape condition at $y =0$, whereas the ensemble of ICs for the BM was set much closer to its respective escape condition. This discrepancy is due to the fact that the standard phase space configuration for the UM is more robust to changes in the perturbation strength, compared to the phase spaces for the BM. Nevertheless, investigating the transient behaviour of escape field lines originating from a region closer to the highly confined magnetic fields (modelled by $y > 0.5$ on the UM) may have important implications for further understanding of plasma-wall interactions in tokamaks.

\begin{figure}[h!]
\centering
\includegraphics[scale=1.0]{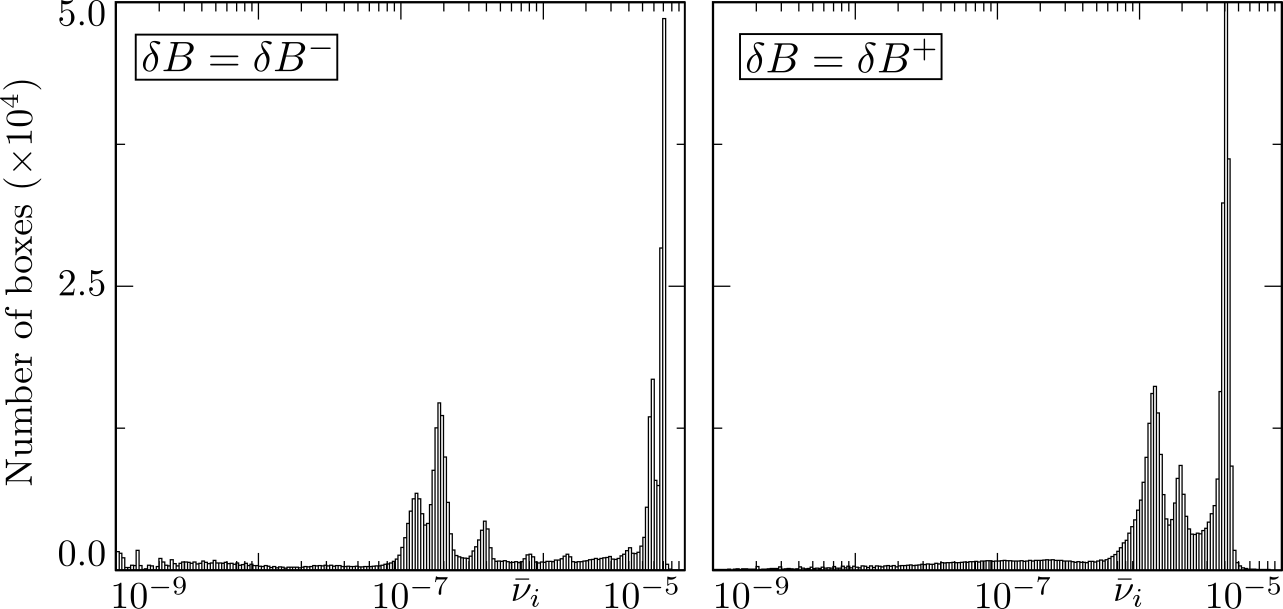}
\caption{Histogram distributions of the mean transient measure $\langle \nu_i \rangle$ for the escape trajectories of the UM, considering both $\delta B=\delta B^{-}=1.334\%$ (left) and $\delta B=\delta B^{+}=1.836\%$ (right). All parameters were kept the same as for Fig.\ \ref{fig:TMA_um_profiles}.}
\label{fig:TMA_um_histograms}
\end{figure}

As a complementary statistical result to Fig.\ \ref{fig:TMA_um_profiles}, we show in Fig.\ \ref{fig:TMA_um_histograms} the computed histogram distributions associated with the $\langle \nu_i \rangle$ profiles.

Figure \ref{fig:TMA_um_histograms} displays rather different distributions in comparison to Fig.\ \ref{fig:TMA_bm_histograms} for the BM. This is expected since for the UM we analyse ICs placed far from the escape condition. We note that the average behaviour of escaping trajectories has similar high peaks around $10^{-5}$, depicted by dark red colours in both panels of Fig.\ \ref{fig:TMA_um_profiles} that are related to the area nearest to the ensemble of ICs. However, other minor peaks are in different positions on the $\langle \nu_i \rangle$ range; While $\delta B^{-}$ presents secondary peaks far from the first one, around $10^{-7}$, the secondary peaks of the distribution for $\delta B^{+}$ are closer to the primary. One way to interpret this result is that the average behaviour of the escaping orbits for $\delta B^{-}$ is more influenced by regions depicted in dark green ($\langle \nu_i \rangle \approx 10^{-7}$), while for $\delta B^{+}$ the regions of influence are closer to the area where the ensemble of ICs was set, namely the regions depicted in red and yellow on the right panel of Fig.\ \ref{fig:TMA_um_profiles}. 

\begin{figure}[h!]
\centering
\includegraphics[scale=1.0]{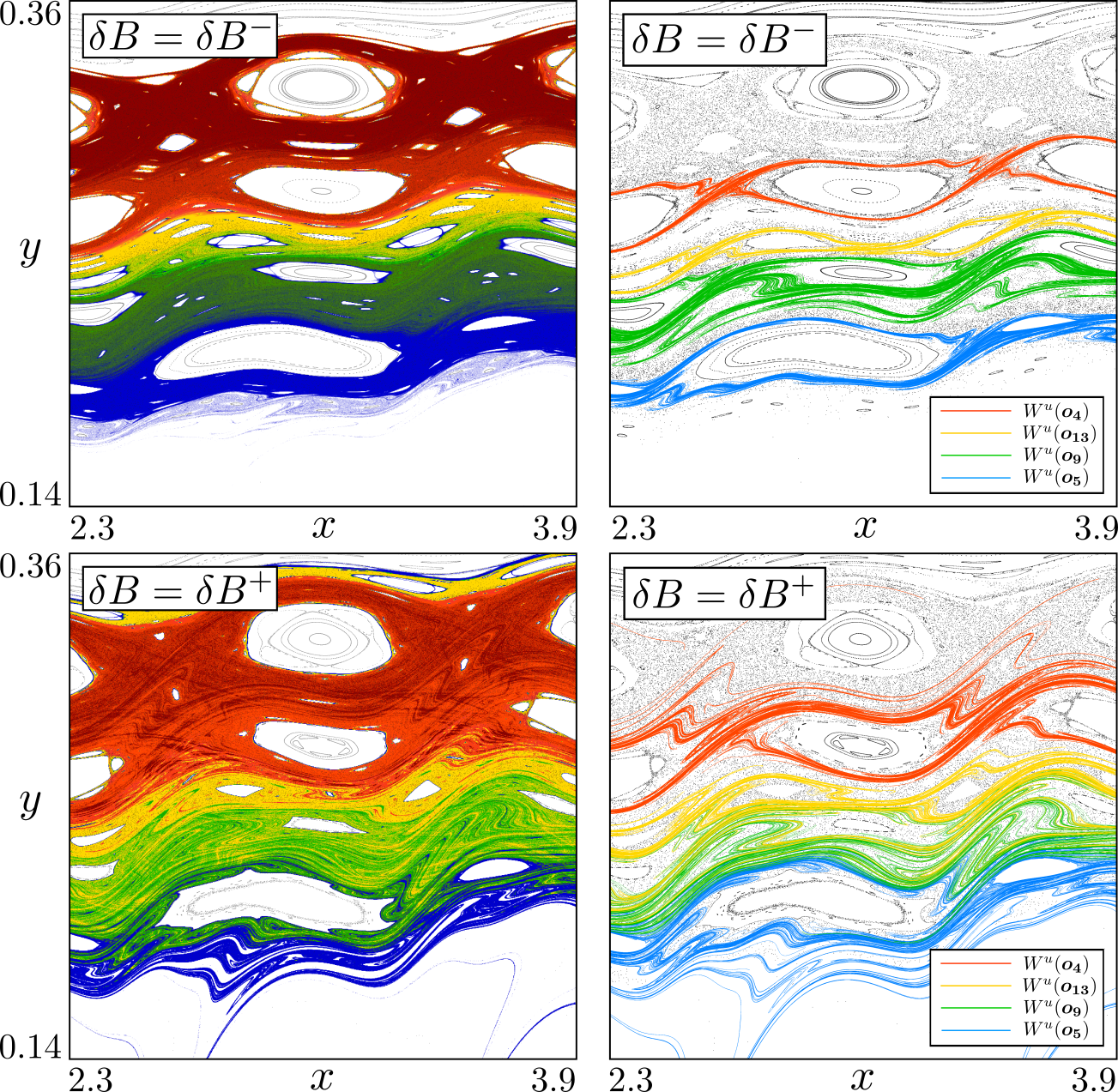}
\caption{Comparison between the $\langle \nu_i \rangle$ profile (left panels) and unstable invariant manifolds $W^{u}$ (right panels) associated with four different chains of islands. Their respective phase spaces are depicted in grey on the background. All parameters and the colour scale (omitted) are the same as in Fig.\ \ref{fig:TMA_um_profiles}.}
\label{fig:TMA_um_mani}
\end{figure}

Moreover, Fig.\ \ref{fig:TMA_um_mani} portrays the last result related to the transient analysis for the UM. We present, in a similar fashion to what was shown for the BM, the comparison between the uncovered structural details and the relevant invariant manifolds over an amplified region of the phase space. It was set $x \in [2.3, 3.9]$ and $y \in [0,14,0.36]$ to make the highlighted structures more evident and, differently from the analysis for the BM, we trace only the unstable manifolds associated with four distinct chains of islands present in the phase space of the UM. Tracing the correspondingly stable manifolds altogether would impair the visualisation, undermining the proposed visual comparison. 

Figure \ref{fig:TMA_um_mani} stress the straightforward link between the average behaviour of all escaping orbits and the underlining invariant manifolds associated with the UM dynamics. Initially, for both values of $\delta B$, the upper region is heavily occupied, as expected for areas closer to the ICs. However, due to the different phase space configurations, the path experienced by all escaping orbits differs in each case; For $\delta B^-$ regions depicted by the colour gradient are well-defined and separated, while for $\delta B^+$ we easily observe a stronger mixing of colours, especially for green and yellow ($\langle \nu_i \rangle \approx 10^7 ~ \text{to} ~ 10^6$). This difference is also revealed by the shapes of all traced unstable manifolds that, for $\delta B^-$, are significantly more restrained in comparison to $\delta B^+$. The strong mixing of the erratic unstable manifolds outlined in the last panel of Fig.\ \ref{fig:TMA_um_mani} might be a fitting explanation of why $\delta B^+$ is a value for the perturbation strength that enhances the escape of this system. 
    
As a final general comment, it is worth paying close attention to all highlighted areas surrounding islands in both systems. The first panel of Fig. \ref{fig:TMA_bm_mani} provides an explicit example for the BM, while the first panel of Fig. \ref{fig:TMA_um_mani} highlights the closer neighbourhood of the upper chain of islands in the UM with a red colour gradient. This observation is also closely related to stickiness phenomena.

\section{Conclusions}
\label{conclusion}
In this work, we investigate differences between distinct magnetic configurations, represented by the phase spaces of two symplectic maps, namely Boozer and Ullmann maps. Our study initially relies on an escape analysis that determines values for the models' perturbation strengths which produces magnetic configurations that either enhance or restrict escaping field lines. Once these values are determined, we employ two different numerical methods, developed to study important aspects of mixed phase spaces, to compare and illustrate the general behaviour of open field lines in these two different magnetic configurations.

First, the recurrence-based approach proved to be suitable for detecting stickiness in both models. The numerical results shown in Sec.\ \ref{results} compare the presence of sticky trajectories considering the aforementioned magnetic configurations. Based on these results, it is possible to infer that the stickiness phenomena is relatively more noticeable on magnetic configurations that restrict escaping field lines. In addition, the method allows an efficient detection of trajectories that widely differs from the average chaotic behaviour. This result has important implications while considering that, due to the magnetic configuration, escaping particles from the plasma may access additional confinement regions in the nearest surroundings of magnetic islands, specifically at the plasma edge.   

The second analysis, based on a practical numerical method that visually illustrates distinct transient behaviour of escaping field lines, shows how the spatial organisation of relevant invariant manifolds can be linked to the average dynamical evolution considering different magnetic configurations on both models. For configurations that enhance the escape, the computed invariant manifolds present much more erratic and intertwined, creating notable transport channels that are absent when compared to the behaviour of the manifolds for configurations that restrict escaping field lines. These results suggest that the manifolds' complex organisation may construct suitable transport channels that can exhaust unwanted particles from the plasma edge in a controlled manner, preventing the thermal load in locations at the inner wall that are not prepared for extracting the high heat flux from the plasma. 

The two selected methods explore chaotic trajectories that either experience stickiness, or are constantly influenced by underlying invariant manifolds present in the mixed phase spaces. These two important aspects are inherently connected to transport and diffusion properties that change depending on the given magnetic configuration. These analyses may improve our understanding of the general behaviour of magnetic field lines that confines fusion plasma in tokamaks.     

\section*{Acknowledgements}
M. S. Palmero was supported by the São Paulo Research Foundation (FAPESP) from Brazil, under Grants Nos. 2018/03000-5 and 2020/12478-6. Iberê L. Caldas acknowledges the Conselho Nacional de Desenvolvimento Científico e Tecnologico (CNPq) from Brazil, under Grant Nos. 407299/2018-1, 302665/2017-0, and the São Paulo Research Foundation (FAPESP) from Brazil under Grant No. 2018/03211-6.








\bibliographystyle{elsarticle-num} 
\bibliography{confine_escape}


\end{document}